\documentclass[a4paper,11pt]{article}
\pdfoutput=1 
\usepackage{jcappub} 
\usepackage{lineno}
\usepackage[T1]{fontenc} 
\usepackage{multirow}
\usepackage{longtable}
\usepackage{tabularx}
\usepackage{booktabs}                      
\usepackage{ifthen}
\usepackage{siunitx}
\usepackage{xcolor}
\usepackage{hyperref}
\usepackage{url}
\usepackage{mathrsfs}
\usepackage{amsmath, amssymb}
\usepackage{caption}
\usepackage{subcaption} 
\usepackage{titlesec}
\usepackage{array}
\usepackage{float}
\usepackage{afterpage}
\usepackage[normalem]{ulem}
\newcolumntype{?}[1]{!{\vrule width #1}}

\arxivnumber{2501.02538}
\title{\boldmath Global 21 cm signal: 
a promising probe of primordial features}

\author[a,1]{Suvedha Suresh Naik\note{Corresponding author.}}
\author[a, b]{Pravabati Chingangbam,}
\author[c]{Saurabh Singh,}
\author[d]{Andrei Mesinger,}
\author[e]{Kazuyuki Furuuchi}

\affiliation[a]{Indian Institute of Astrophysics,
	Koramangala II Block, Bangalore 560 034, India}
\affiliation[b]{Korea Institute for Advanced Study (KIAS), 85 Hoegiro, Dongdaemun-gu, Seoul, Republic of Korea-02455}
 \affiliation[c]{Raman Research Institute, CV
Raman Avenue, Sadashivanagar, Bangalore, 560 080, India}
 
\affiliation[d]{Scuola Normale Superiore, Piazza dei Cavalieri 7, I-56125 Pisa, Italy}

 \affiliation[e]{Manipal Centre for Natural Sciences,
 Manipal Academy of Higher Education,
 Manipal 576 104, Karnataka, India}
 
\emailAdd{suvedha.naik@iiap.res.in}
\emailAdd{prava@iiap.res.in}
\emailAdd{saurabhs@rri.res.in}
\emailAdd{andrei.mesinger@sns.it}
\emailAdd{kazuyuki.furuuchi@manipal.edu}

\abstract{ 
Inflationary models that involve bursts of particle production generate bump-like features in the 
primordial power spectrum of density perturbations. 
These features influence the evolution of density fluctuations, 
leaving their unique signatures in
cosmological observations. 
A detailed investigation of such signatures would help constrain physical processes during inflation. 
With this motivation, 
{the goal of this paper is two-fold. First,} we conduct a detailed analysis of the effects of bump-like primordial features on 
the sky-averaged 21 cm signal. 
Using semi-numerical simulations, 
we demonstrate that the primordial features can 
significantly alter the ionization history and 
the global 21 cm profile, 
making them a promising probe of inflationary models. 
We found a special scale (namely, the turnover wavenumber, $k^{\rm turn}$) 
at which the effect of primordial 
bump-like features on the global 21 cm profile vanishes.
Also, we found that the behaviour of the primordial features 
on the global profile and ionization history 
are quite opposite for $k > k^{\rm turn}$ and $k < k^{\rm turn}$.
We {trace the root cause of} these behaviours 
to the effects of primordial features on 
the halo mass function at high redshifts.
Furthermore, we discuss the degeneracy 
between the astrophysical parameters and the primordial features in detail. 
Secondly, for a fixed set of astrophysical parameters, we derive upper limits on the amplitude of bump-like features in the range $10^{-1} < k\,[{\rm Mpc}^{-1}] < 10^2$ using current limits on optical depth to reionization from CMB data by {\it Planck}.
}
\begin{document}
\maketitle
\flushbottom
\section{Introduction}
\label{sec:intro}
%
Current observations of the Cosmic Microwave Background (CMB) and 
Large Scale Structures (LSS) are well explained by 
the theory of cosmic inflation \cite{Guth:1981,LINDE:1982,Albrecht:1982, 
STAROBINSKY198099,stato:1981,kazanas:1980}. According to the theory 
of inflation, primordial density perturbations originated as quantum 
fluctuations during inflation and evolved to form the structures in the 
universe. Therefore, cosmological observations are expected to carry 
imprints of the physical processes that occurred during inflation. 
Judicious extraction of this primordial information can then help to 
constrain models of inflation.

Despite the broad consistency between current observations and 
the $\Lambda$CDM model, characterized by a nearly scale-invariant 
primordial power spectrum, the data suggest possible deviations from 
scale-invariant behavior at specific spatial scales. 
Several classes of inflationary models predict 
scale-dependent features in the primordial power spectrum, 
commonly referred to as "primordial features" 
(see, e.g., \cite{Akrami:2018odb, Chluba:2015bqa, Beutler:2019ojk, Barnaby:2009dd, Furuuchi:2020klq, Jain:2008dw} and references therein). 
Testing these models requires comparing their predictions with 
observations across different wavenumber ranges. Over the past two 
decades, several studies have sought to constrain primordial features 
within the wavenumber range $10^{-4} < k\,[{\rm Mpc}^{-1}] < 0.2$, using 
CMB and LSS observations (see, e.g., \cite{Akrami:2018odb,Beutler:2019ojk} 
for the latest results). The smaller scales corresponding to 
$1 < k\,[{\rm Mpc}^{-1}] < 10$ are probed by the Lyman-$\alpha$ forest 
(see, e.g., \cite{Irsic:2017yje} for constraints on dark matter models) 
and weak-lensing measurements 
\cite{sdss14:2018ApJS..235...42A,DES:2017qwj}. Scales beyond this region 
remain largely unconstrained at present.

In this work we focus on a class of 
theoretically well-motivated inflationary models 
involving bursts of particle production during inflation 
that predict bump-like features in the primordial power spectrum 
\cite{Chung:1999ve,Barnaby:2009mc,Barnaby:2009dd,Pearce:2017bdc,Furuuchi:2015foh,Furuuchi:2020klq,Furuuchi:2020ery}. 
Such bursts of particle production may occur naturally in inflation models 
based on higher-dimensional gauge theories 
\cite{Furuuchi:2015foh,Furuuchi:2020klq,Furuuchi:2020ery}. 
This strongly motivates searches for signatures of 
bump-like features in cosmological observations\footnote{The bump-like primordial 
features can also be expected from different theoretical motivations besides 
particle production during inflation. 
For example, models involving axion isocurvature perturbations 
that lead to an enhanced isocurvature power spectrum can also imprint such features \cite{Chung:2016wvv}. Refs.~\cite{Sakharov:1994id,Khlopov:1998uj} discuss inhomogeneities in the initial perturbations 
of axionic cold dark matter that may also give rise to bump-like features.
}. 
The signatures of bump-like features have been investigated 
using CMB observations \cite{Naik:2022mxn} and galaxy two-point 
correlation functions \cite{Ballardini:2022wzu}. 
Currently available data provide upper limits on the amplitudes of these 
features in the range $10^{-4} < k\,[{\rm Mpc}^{-1}] < 0.2$.

The 21~cm signal from the hyperfine transition of neutral hydrogen 
is expected to revolutionize our understanding of the physics 
of the early universe. 
The redshifted 21 cm signal will provide access to 
wavenumbers beyond those probed by current observations. 
Moreover, the tomographic nature of the signal will yield 
vast datasets to test theories governing inflation, cosmic dawn, and 
the Epoch of Reionization (EoR). 
Several ongoing and upcoming experiments aim to measure the 
power spectrum of the 21 cm signal at high redshifts, such as 
SKA \cite{ska}, 
HERA \cite{DeBoer:2016tnn}, 
LEDA \cite{LEDA:2018MNRAS}, 
LOFAR - 2.0 \cite{LOFAR2:2021A&A...652A..37E}, 
MWA Phase II \cite{MWA2:2018PASA...35...33W}, 
NenuFAR \cite{NenuFAR:2017arXiv171206950A}, etc. 
Besides the power spectrum, 
experiments like 
EDGES \cite{EDGESBowman:2007su}, SARAS \cite{SARAS1:2013ExA....36..319P, SARAS2:Singh:2017syr}, 
BIGHORNS \cite{BIGHORNS:2015PASA}, 
PRIZM \cite{PRIZM:2019JAI}, 
REACH \cite{REACH:2022NatAs...6.1332D}, 
MIST \cite{MIST:2024MNRAS.530.4125M}, 
RHINO \cite{RHINO:2024arXiv241000076B}, etc., 
aim to measure the sky-averaged 21 cm signal 
from the cosmic dawn to the reionization epoch. 
In addition, there are several space-based proposals for studying 
the dark ages and cosmic dawn, such as 
ALO \cite{ALO:2024AAS...24326401K}, 
ROLSES \cite{ROLSES:2021PSJ.....2...44B}, 
LuSEE-night \cite{LuseeNight:2023arXiv230110345B}, 
PRATUSH \cite{Pratush:2023ExA....56..741S}, etc. 
The profile of the global 21 cm signal 
holds crucial information regarding the formation of the first 
stars and galaxies, 
heating of the IGM by X-ray photons, and 
the reionization of neutral hydrogen by UV photons 
and other physical processes \cite{Pritchard:2011xb}. 
Since the 21~cm signal from the high-redshift universe encodes the 
signatures of density fluctuations and astrophysical processes, 
it can serve as an important probe of inflationary models. 
In a recent work by \cite{Yoshiura:2019zxq}, the impact of a modified primordial power spectrum on the averaged 21 cm signal was studied in detail. The primordial features with a dip in the power spectrum were investigated in the context of the 21~cm power spectrum and bispectrum in ref.\cite{Balaji:2022zur}. Ref.~\cite{Munoz:2019hjh} studied in detail the potential of the 21 cm signal to probe the matter power spectrum on small scale.

In a previous work by some of the authors \cite{Naik:2022wej}, 
it was demonstrated that the 21 cm power spectrum, 
spanning multiple redshifts combined with expected noise from SKA-Low, 
has the potential to probe bump-like features 
within the range $0.1 \le k\,[{\rm Mpc}^{-1}] \le 1.0$. 
In this work, we extend the previous analysis and focus on how 
the sky-averaged 21~cm signal can be used as a potential probe 
of bump-like features over a wider range of wavenumbers. 
We conduct a systematic analysis to examine the impact 
of bump-like primordial features on the global 21 cm signal 
using semi-numerical simulations. 
The presence of bump-like features in the primordial power spectrum 
enhances correlations in density fluctuations. 
Consequently, these features can significantly modify 
the reionization history and the global 21 cm profile, 
allowing us to investigate their signatures 
in the range $10^{-1} \lesssim k\,[{\rm Mpc}^{-1}] \lesssim 10^2$. 
We find that the primary factor contributing to this modification 
is the change in the halo mass function, 
or the number density of halos, 
due to the addition of primordial features. 
Interestingly, we identify a specific scale 
where the effects of primordial features are nullified, 
with their impact on the global 21 cm profile reversing 
above and below this scale. 
The distinct signatures of bump-like features on the evolution 
of the neutral fraction of hydrogen and the global 21 cm profile 
make them a promising probe of primordial features. 
Furthermore, the impact of primordial features on the 
reionization history enabled us to place upper limits on 
the amplitude of these features 
in the range $10^{-1} \lesssim k\,[{\rm Mpc}^{-1}] \lesssim 10^2$, 
for a given set of astrophysical parameters, 
using {\it Planck}'s measurement of 
the optical depth to reionization. 

The remainder of the paper is organised as follows: 
section~\ref{sec:model} briefly discusses the parameterization 
of inflationary models that predict bump-like features. 
In section~\ref{sec:global21}, we provide a brief overview 
of the global 21 cm profile and the details of our simulations. 
Section~\ref{sec:effects} presents our main results, 
showing the effects of primordial features on the ionization history 
and global 21~cm profiles, along with a detailed analysis 
to understand these effects. 
In section~\ref{sec:s5}, we compare these effects with those 
arising from variations in the astrophysical parameters 
of the first stars and galaxies.  
Section~\ref{subsec:constraints} discusses the observational constraints 
on bump-like features obtained using the ionization history. 
Finally, in section~\ref{sec:Discussion}, we summarise our results 
and comment on future directions. 
 
    %
\section{Primordial features due to particle productions 
during inflation}
\label{sec:model}
The primordial scalar power spectrum assumed in the 
concordance $\Lambda$CDM model is 
given by 
\begin{equation}\label{eq:PS_PowerLaw}
P_S
=
A_s \left(\frac{k}{k_\ast}\right)^{n_s-1}\,,
\end{equation}
where the pivot scale is typically chosen as 
$k_\ast = 0.05~\text{Mpc}^{-1}$, 
and $A_s$ and $n_s$ parameterize the 
amplitude and spectral tilt of the power spectrum, respectively. 
From {\it Planck} data, we obtain 
$\ln (10^{10} A_s) = 3.044 \pm 0.014$ and 
$n_s = 0.9649 \pm 0.0042$ 
at $68\%$ confidence level \cite{Akrami:2018odb}.

We study a class of inflation models 
\cite{Chung:1999ve,Barnaby:2009mc,Barnaby:2009dd,Pearce:2017bdc,Furuuchi:2015foh,Furuuchi:2020klq,Furuuchi:2020ery} 
in which the inflaton field $\phi$ is coupled to a real 
scalar field $\chi$ through the interaction term 
\begin{equation}
g^2 
(\phi - \phi_0)^2 \chi^2 \,,
\label{eq:g_square}
\end{equation}
where $g$ is the dimensionless coupling constant. 
When the inflaton field value crosses $\phi = \phi_0$, 
a burst of $\chi$ particle production occurs as they 
become instantaneously massless. 
When such an event occurs 
during the observable range of $e$-folds of inflation, 
it manifests in the primordial power spectrum as a 
bump-like feature.

\begin{figure}[tbp]
	\centering             
    \includegraphics[width=0.6\textwidth]{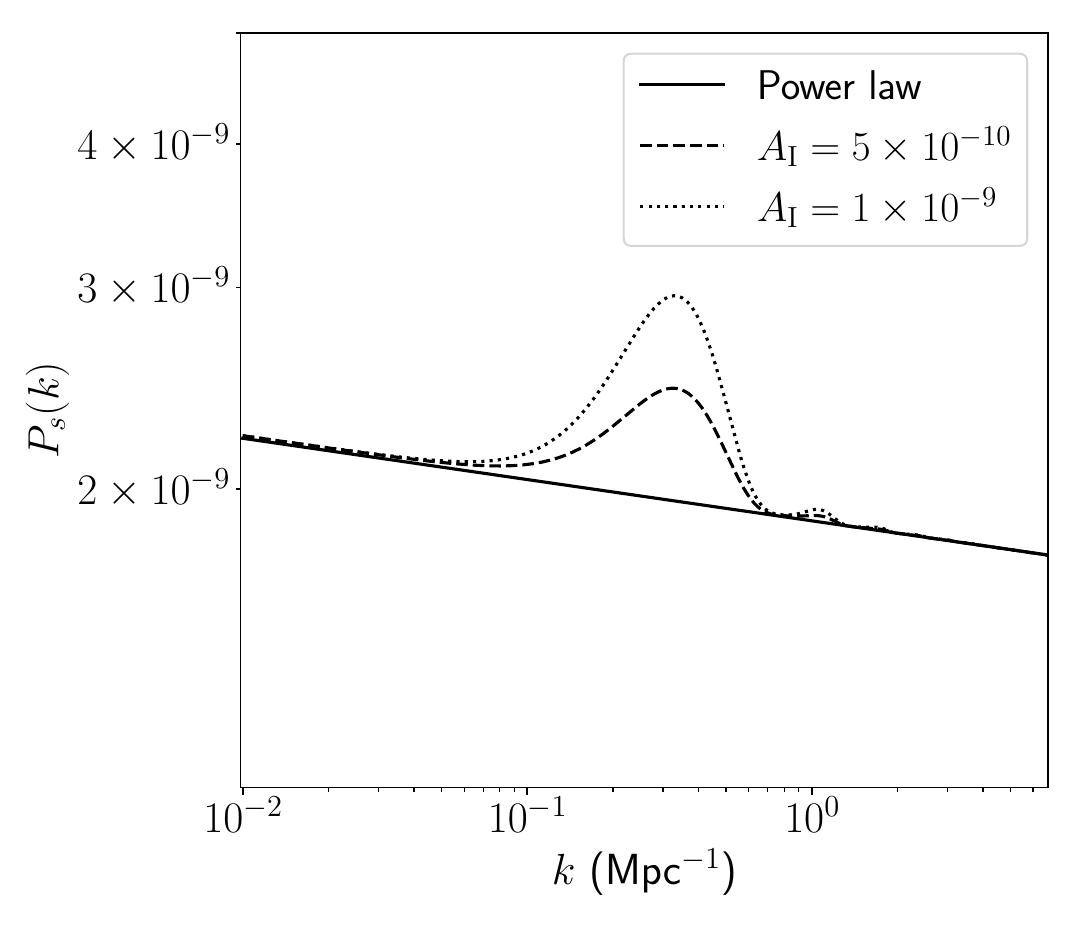}
	\caption{The primordial power spectrum with a bump-like feature at $k_{\rm peak} = 0.3\,{\rm Mpc}^{-1}$ is shown by dashed and dotted curves for 
    $A_{\rm I}=1.0\times10^{-9}$ and
	$5.0\times10^{-10}$, respectively.
	The primordial power spectrum for the power law 
    as in eq.~\eqref{eq:PS_PowerLaw} is plotted
	as a solid line.
    }
    \label{fig:singlebump}
\end{figure}
We parameterize the 
primordial power spectrum with bump-like 
features by including the dominant and subdominant
contributions to the power spectrum, 
calculated analytically with one-loop approximations \cite{Pearce:2017bdc}:
\begin{equation}
P_S
=
A_s \left(\frac{k}{k_\ast}\right)^{n_s-1}
+
A_{\rm I} \sum_i \left(\frac{f_1(x_i)}{f_1^{\rm max}}\right)
+
A_{\rm II} \sum_i \left(\frac{f_2(x_i)}{f_2^{\rm max}}\right)\,,
\label{eq:PS_bump_ch4}
\end{equation}

where 
$A_{\rm I}$ and $A_{\rm II}$
are amplitudes of dominant and
subdominant contributions, respectively. 
The amplitudes depend on the coupling parameter as 
\begin{align}
A_{\rm I} &\simeq 6.6\times 10^{-7} g^{7/2} \,,\label{eq:A1}\\
A_{\rm II} &\simeq 1.1 \times 10^{-10} g^{5/2} \ln\left(\frac{g}{0.0003}\right)^2\,. \label{eq:A2}
\end{align}
The scale dependence of the contributions is given by the 
dimensionless functions
\begin{align}
f_1(x_i) &\equiv
\frac{\left[\sin(x_i)-{\rm SinIntegral}(x_i)\right]^2}{x_i^3} \,,\label{eq:f1}\\
f_2(x_i) &\equiv
\frac{-2x_i\cos(2x_i)+(1-x_i^2)\sin(2x_i)}{x_i^3}\,, \label{eq:f2}
\end{align}
where 
$x_i \equiv \frac{k}{k_i}$.
The parameter
$k_i\,[{\rm Mpc}^{-1}]$
is related to the position of the $i^{\rm th}$ feature on the primordial
power spectrum, which 
peaks around $3.35\times  k_i$, and we define
\begin{equation}
k_{{\rm peak},i} = 3.35\times  k_i\,. 
\label{eq:kpeak}
\end{equation}

For simplicity, in this paper, we focus on the case of a single 
episode of particle production during inflation, 
leading to a single bump-like feature on the 
primordial power spectrum with amplitude\footnote{The subdominant amplitude $A_{\rm II}$ is not an independent parameter and can be written as 
$A_{\rm II} 
\simeq
(2.9\times 10^{-6})
A_{\rm I}^{5/7}
\left[ 
\ln A_{\rm I}^{4/7} + 24
\right]$.} $A_{\rm I}$
and peak location $k_{\rm peak}\, [{\rm Mpc}^{-1}]$.
Figure~\ref{fig:singlebump} shows a comparison of such a power spectrum with the nearly scale-invariant form.

\section{The global 21~cm profile}
\label{sec:global21}

In this section, we provide a brief overview of the 
global 21 cm profile
followed by the details 
of the simulations employed in this study.
Theoretical models and numerical simulations based on radiative processes in the high-redshift universe 
have provided predictions for the expected 
evolution of the sky-averaged 21 cm signal, spanning from the dark ages to the completion of reionization 
(see, e.g., \cite{Furlanetto:2006jb, Furlanetto:2009astro2010S..82F, Pritchard:2012RPPh...75h6901P, Mesinger:2019cosm.book}
for comprehensive reviews).

%
In the Rayleigh-Jeans approximation, 
the specific intensity of 21 cm radiation originating from neutral hydrogen (HI) 
can be expressed in terms of its brightness temperature ($T_b$). 
The differential brightness temperature $\delta T_b$, 
which captures the contrast between the spin temperature $T_S$ and 
background radiation $T_\gamma$, 
is given by
(e.g., \cite{Madau:1996cs, Barkana:2000fd, Furlanetto:2006jb}):
\begin{equation}
    \delta T_{b} (z) \approx
    27\, x_{\rm HI}\, 
    (1+ \delta_{\rm nl}) \, 
    \left(\frac{H(z)}{{dv_r/dr} + H(z)}\right)
    \left(1 - \frac{T_\gamma}{T_S}\right)
    \left(\frac{1+z}{10} \frac{0.15}{\Omega_m h^2}\right)^{1/2}
    \left(\frac{\Omega_b h^2}{0.023}\right)\,
    {\rm mK}\,,
    \label{eq:Tb}
\end{equation}
where 
$\delta_{\rm nl}$ represents the fractional baryon density perturbation, 
$x_{\mathrm{HI}}$ denotes the neutral hydrogen fraction, 
$H(z)$ is the Hubble parameter, 
and ${dv_r}/{dr}$ is the 
line-of-sight component of the proper velocity gradient. 
The spin temperature, $T_{\rm S}$, describes the relative population 
of atoms in the two hyperfine levels. 
Thus, the observable quantity - the differential brightness temperature of the 21 cm signal
- is shaped by the underlying density fluctuations, 
as well as the physical processes driving the formation of the earliest luminous sources 
and the reionization of the universe.

\begin{figure}[tbp]
    \centering
    \includegraphics[width=\textwidth]{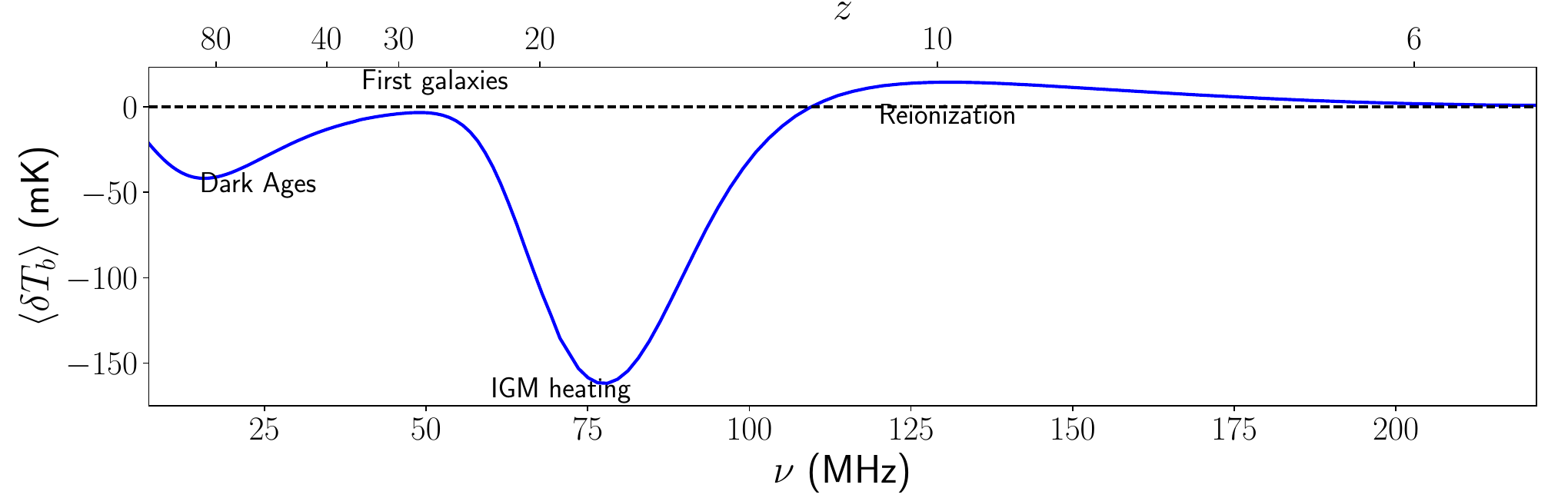}
    \caption{The time evolution of the sky-averaged 21~cm signal predicted from theoretical models 
    and simulated with \texttt{21cmFAST} \cite{Mesinger:2011}.}
    \label{fig:global21_ideal}
\end{figure}

The simplest fundamental quantity used to characterize the 21 cm signal 
is the sky-averaged 21 cm brightness temperature, denoted by $\left<{\delta T_b}\right>$, 
and its evolution across different redshifts, i.e., the global 21 cm profile.
The evolution of $\left<{\delta T_b}\right>$ obtained from a simulation of a representative theoretical model, 
is shown in figure~\ref{fig:global21_ideal}, with some significant physical stages highlighted.
%
%
The global profile includes absorption and emission features, whose 
amplitude and timing depend significantly on the underlying astrophysical processes 
\cite{Madau:1996cs,Barkana:2000fd,Furlanetto:2006jb}. 
The initial density fluctuations
are influenced by cosmology through the primordial power spectrum, while 
the evolution of the quantities $\delta_{\rm nl}$, $x_{\rm HI}$, and $T_S$
is determined by astrophysical processes.

\subsection{Semi-numerical Simulations of the 21 cm Signal}
\label{subsec:21cm_sim}
We utilize \texttt{21cmFASTv3}\footnote{\url{https://github.com/21cmfast/21cmFAST}} 
\cite{Mesinger:2011,Murray:2020trn}, a semi-numerical simulation code, 
to model the cosmological 21 cm signal. 
\texttt{21cmFAST} employs simplified physical modeling to 
efficiently generate realizations of the density, ionization, spin temperature fields, 
and velocity gradients, which are then combined to calculate the brightness temperature 
of the 21 cm signal using eq.~\eqref{eq:Tb}.

We adopt the Friedmann-Lemaître-Robertson-Walker cosmological framework 
with flat spatial geometry. 
The concordance $\Lambda$CDM model is defined by several key parameters: 
the baryon density $\omega_b = \Omega_b h^2$, 
the cold dark matter density $\omega_{\text{cdm}} = \Omega_{\text{cdm}} h^2$, 
the present Hubble parameter $H_0$, 
the optical depth to reionization $\tau_e$, 
the amplitude of scalar perturbations $A_{\rm s}$ 
at the pivot scale $k_\ast = 0.05\,{\rm Mpc}^{-1}$ 
(or alternatively $\sigma_8$, the variance of density perturbations 
within a sphere of radius $8 h^{-1}\,{\rm Mpc}$), 
and the spectral index $n_s$.

As mentioned previously, the brightness temperature of the 21 cm signal is affected by 
astrophysical processes, which are modeled through several parameters in \texttt{21cmFAST}.
Those discussed frequently in this paper are: 
\begin{itemize}
	\item $M_{\rm min}$ - 
	the halo mass below which the abundance of
	active star-forming galaxies is exponentially suppressed.  
        $M_{\rm min}$ (in solar mass units $M_\odot$)
        is often expressed in terms of the {virial temperature}
$T_{\rm vir}$ (K) as \cite{Barkana:2000fd} 
\begin{equation}
M_{\rm min}
=
10^8 h^{-1}
\left(\frac{\Omega_m}{\Omega_{m,\,z}} \frac{\Delta_{c,\,z}}{18\pi^2}\right)^{-1/2}
\left(\frac{T_{\rm vir} (\rm K)}{1.98\times 10^4{\rm K}}\right)^{3/2}
\left(\frac{1+z}{10}\right)^{-3/2}
{ M_\odot}\,,
\label{eq:Mmin}
\end{equation}
where 
\begin{align*}
\Omega_{m,\,z} &= \Omega_m \left[\frac{(1+z)^3}{\Omega_m(1+z)^3+\Omega_\Lambda}\right] \,, \\
\Delta_{c,\,z} &=18\pi^2-39{d_z}^2+82d_z\,,
\end{align*}
where $d_z=\Omega_{m,\,z}-1$.
        \item $\zeta$ - the ionizing efficiency of high-$z$ galaxies
        modelled as 
\begin{equation}
\zeta =  30 
\left(\frac{f_{\rm esc}}{0.12}\right)
\left(\frac{f_\ast}{0.05}\right)
\left(\frac{N_\gamma}{4000}\right)
\left(\frac{1.5}{1+n_{\rm rec}}\right)\,,
\label{eq:zeta}
\end{equation}
where $f_{\rm esc}$ is the fraction of ionizing photons that escape into the IGM, 
$f_\ast$ is the fraction of galactic gas into stars,
$N_\gamma$ quantifies the ionizing photons per baryons
and 
$n_{\rm rec}$ gives the typical number of recombination.
 	\item $L_{X<2{\rm keV}}/$SFR - 	the normalization of the soft-band X-ray
	luminosity per unit star formation computed over the band 2 keV.
%
\end{itemize}
Other parameters that are used for modelling the reionization in \texttt{21cmFAST} are:
$\alpha_{\rm esc}$ - the power-law index of $f_{\rm esc}$, 
$\alpha_\ast$ - the power-law scaling of $f_\ast$ with halo mass, 
$t_\ast$ - the star formation time-scale taken as a fraction of the Hubble time, and 
$E_0$ - the lowest energy of the X-ray photon that can escape the galaxy.
\vskip 2mm
\noindent{\bf Fiducial model:}
We define our base model, or the {\em fiducial model}, as follows: the 
concordance $\Lambda$CDM parameters are fixed using the 
best-fit values from the \textit{Planck} 2018 results \cite{Akrami:2018odb}\footnote{The best-fit values obtained 
	by combining data from temperature, polarization, and lensing.}:
$\Omega_b h^2 = 0.022$,
$\Omega_{\text{cdm}} h^2 = 0.120$,
$h = 0.6736$,
$\tau_e = 0.058$,
$\sigma_8 = 0.811$ and
$n_s = 0.965$. 
We set the astrophysical parameters according to \cite{Park:2018ljd}:
$M_{\rm min} = 5\times 10^8 M_\odot$ or $\log T_{\rm vir} = 4.69897$,
$\zeta = 30$ (i.e., $f_{\rm esc} = 0.1$, $f_\ast = 0.05$, and 
$N_\gamma = 5000$),
$\alpha_\ast = 0.5$,
$\alpha_{\rm esc} = -0.5$,
$t_\ast = 0.5$,
$E_0 = 0.5$ {keV}, and 
$L_{X<2{\rm {keV}}}/$SFR $ = 10^{40.5}~\text{erg s}^{-1} M_\odot^{-1} {\rm yr}$.
In \texttt{21cmFAST}, 
the matter power spectrum is calculated using the 
default power-law form \cite{Eisenstein:1997ik,Eisenstein:1997jh}.

In all our simulations, we select a simulation box length of $300$ Mpc, 
beginning with a higher-resolution box containing 
$300$ cells, which is then downsampled to a lower-resolution box with 
$100$ cells. 
This configuration yields a resolution of $3$ Mpc. 
The selected box length and cell size provide a compromise 
between the typical resolution from observations and feasible computational time.
We simulate the brightness temperature across the redshift range 
$5 \lesssim z \lesssim 35$, incorporating calculations of the spin temperature. 


\section{Global 21~cm signal as a probe of bump-like primordial features with fixed astrophysical parameters } 
\label{sec:effects}

As described in section~\ref{sec:model}, 
we parameterize the bump-like primordial features 
by the amplitude of the bump, $A_{\rm I}$, 
and the location of the peak of the bump denoted as $k_{\rm peak}$.  
The theoretically estimated upper limit on the 
coupling parameter is given by 
$g^2 \lesssim 3$ \cite{Pearce:2017bdc}, 
which imposes a constraint on $A_{\rm I}$ via 
equation~\eqref{eq:A1}, resulting in 
$A_{\rm I} < 10^{-6}$. 
The parameter $k_{\rm peak}$ is not constrained by theoretical models. 
Since the bump-like primordial features 
are already constrained at the scales observed 
with the CMB in ref.~\cite{Naik:2022mxn}, 
we explore scales beyond $0.1\, {\rm Mpc}^{-1}$.  
Thus, our analysis encompasses models in the parameter ranges: 
$A_{\rm I} \sim [10^{-10}, 10^{-7}]$ and 
$k_{\rm peak} \geq 0.1\, {\rm Mpc}^{-1}$. 
In this section, we keep the parameters of the astrophysical model fixed to the values of the 
``fiducial'' model given in section \ref{subsec:21cm_sim} so as to cleanly investigate the effect of the primordial features.  
We revisit the effects of varying the astrophysical parameters in section~\ref{sec:s5}.

\subsection{Effects on $\left<x_{\rm HI}\right>$ and $\left<\delta T_b\right>$}
\label{sec:s4.1}
\begin{figure}[tbp]
    \centering
    \includegraphics[width=1.\textwidth]{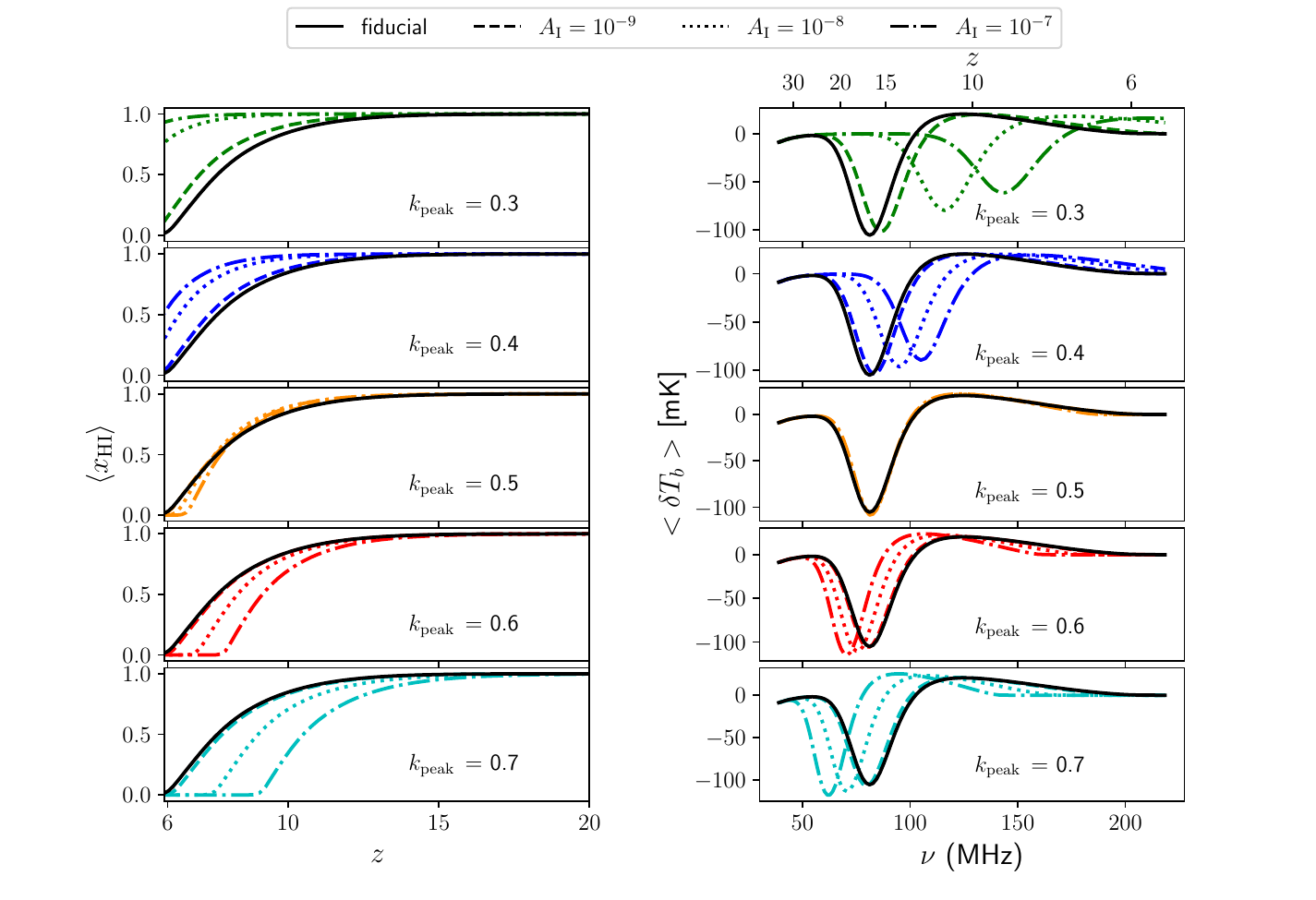}
    \caption{[Left] The ionization histories and [right] the global 21 cm profiles 
for single bump models at different values of $k_{\rm peak}$. 
Each row corresponds to a specific $k_{\rm peak}$ value, 
with variations in the amplitude parameter $A_{\rm I}$ indicated by 
different line styles. The solid black curve represents the fiducial model 
without primordial features.}
    \label{fig:global21_all}
\end{figure}

We begin by examining the effects of bump-like features 
on the averaged neutral hydrogen fraction 
$\left<x_{\rm HI}\right>$ and brightness temperature of the 
21 cm signal $\left<\delta T_b\right>$. 
In figure~\ref{fig:global21_all}, we present 
the redshift evolution of $\left< x_{\rm HI} \right>$ 
and $\left< \delta T_{\rm b} \right>$, 
for various values of $k_{\rm peak}$ (indicated in each panel) 
and amplitudes $A_{\rm I}$ (represented by different line styles).  

%
The impact of bump-like features on both 
$\left< x_{\rm HI} \right>$ 
and $\left< \delta T_{\rm b} \right>$ can be substantial. 
The ionization histories, depicted by the redshift evolution of $\left< x_{\rm HI} \right>$, show that 
$k_{\rm peak}$ affects the timing of 
the completion of reionization. 
As $k_{\rm peak}$ increases, 
its effect transitions from 
late reionization to early reionization, relative to the fiducial model. 
We observe that this turnover of effects of primordial features 
occurs around the scale 
$k \sim 0.5 \, {\rm Mpc}^{-1}$ 
for the current choice of the fiducial model.
We will refer to this scale as the 
``turnover scale'', $k^{\rm turn}$, hereafter. 
We find a similar effect of $k_{\rm peak}$ on $\left< \delta T_{\rm b} \right>$ which also exhibits a transition 
around the same turnover scale. 
This special scale $k^{\rm turn}$ is crucial for our subsequent discussion 
as it marks the division of the effects of bump-like features  
into two distinct wavenumber regimes: 
\begin{itemize}
    \item $\mathbf{k_{\rm peak} < k^{\rm turn}}$:
    As seen in the top two rows of figure \ref{fig:global21_all}, a smaller value of $k_{\rm peak}$ 
    corresponds to a relatively delayed end of reionization compared 
    to the fiducial model. 
    Then for fixed $k_{\rm peak}$, increasing $A_I$ 
    has the effect of increasing the delay of reionization, as seen in the plots of $\left< x_{\rm HI} \right>$. 
    For $\left< \delta T_{\rm b} \right>$ we obtain a systematic shift of the profile towards lower redshifts. This shift is associated with a broadening of the width of the absorption trough and shallowing of the depth. Further, the rate of change with respect to $A_I$ increases with increasing difference of $k_{\rm peak}$ from $k^{\rm turn}$.

    \item $\mathbf{k_{\rm peak} > k^{\rm turn}}$:
    In this regime, as shown by the last two rows of figure \ref{fig:global21_all},  
    we obtain trends in the behaviour of both $\left< x_{\rm HI} \right>$ and  
    $\left< \delta T_{\rm b} \right>$ that are opposite to the above regime.

    \item $\mathbf{k_{\rm peak} \sim k^{\rm turn}}$:
    For this special scale, the curves of both $\left< x_{\rm HI} \right>$ 
    and  $\left< \delta T_{\rm b} \right>$, 
    corresponding to different values of $A_{\rm I}$,  
    are indistinguishable from the fiducial model 
    (see middle panels of both columns of figure \ref{fig:global21_all}). 
    This wavenumber thus represents a critical scale at which the bump-like features 
    exert no significant influence on the 
    profile of the global 21 cm signal.
\end{itemize}

Having discovered the existence of the interesting scale $k^{\rm turn}$, 
we next investigate its physical origin. 
Since $\delta T_{\rm b}$ contains contributions from  
density fluctuations $\delta_{\rm nl}$, spin temperature $T_S$ and 
neutral fraction of hydrogen $x_{\rm HI}$, we examine the effect of the 
bump-like feature on these quantities separately. 
The details are provided in appendix~\ref{app:}.
All of these quantities depend on 
the collapsed fraction, $f_{\rm coll}(z)$ - 
the fraction of gas  
inside collapsed objects at redshift $z$, 
which in turn is derived 
using the halo mass functions. Therefore, understanding the effect of 
primordial features on the halo mass function is key to understanding 
the physical origin of $k^{\rm turn}$. This is what we address in the 
next subsection.

\subsection{The Halo Mass Function in the presence of 
primordial features}
\label{subsubsec:HMF}
%
The Halo Mass Function (HMF) 
denoted as $\frac{dn}{d\ln M}\, [{\rm Mpc}^{-3}]$, 
represents the co-moving number density of halos 
in the mass range $M$ to $(M+dM)$, and 
is given by \cite{Press_Schechter1974ApJ...187..425P}
\begin{equation}
    \frac{dn}{d \ln M} 
    =
    \frac{\rho_m}{M}
    \frac{-d(\ln \sigma)}{dM}
    \nu 
    f(\nu)
    \,,
    \label{eq:dndM}
\end{equation}
where 
$\rho_m$ represents the average matter density at $z=0$ and 
$\nu \equiv  \frac{\delta_c}{D(z) \sigma(M)}$, 
where $\delta_c$ is the critical overdensity, 
$D(z)$ is the growth factor. $\sigma(M)$ is the 
variance of the initial density fluctuation field, which is linearly extrapolated to the present epoch and smoothed with a filter $W(kR)$ of scale $R$. It can be expressed as
\begin{equation}
    \sigma^2(R) =
    \frac{1}{2\pi^2}
    \int_0^\infty 
    dk \,
    k^2 P_m(k)
    W^2(kR)
    \,,
    \label{eq:smoothed_sigma}
\end{equation}
where $P_m(k)$ is the matter power spectrum. 
In \texttt{21cmFAST}, a real space top-hat filter is used by default, 
whose Fourier transform is given by 
\begin{equation}
    W(kR) = 3 \left[\frac{\sin(kR) - (kR) \cos(kR)}{(kR)^3}\right]\,,
    \label{eq:filter}
\end{equation}
where the radius of the top-hat filter, $R$,  
is given by 
\begin{equation}
    R (M) \equiv 
    \left[\left(\frac{3}{4\pi}\right)
    \left(\frac{M}{\rho_m}\right)\right]^{1/3}\,.
    \label{eq:tophat}
\end{equation}
In \texttt{21cmFAST}, the Sheth-Tormen formula \cite{Sheth:2001dp} 
for the HMF is used to 
calculate the number density of halos. 
In this formalism, the term $\nu f(\nu)$ in 
eq.~\eqref{eq:dndM} is given by
\begin{equation}
    \nu f(\nu) = 
    \sqrt{2 \pi}
    A 
    \left( 1 + \frac{1}{\hat{\nu}^{2q}}\right)
    \hat{\nu}
    \exp{\left(\frac{-\hat{\nu}^2}{2}\right)}
    \,,
    \label{eq:ST_term}
\end{equation}
where $\hat{\nu} =  \sqrt{a} \nu$. 
The parameter values $a = 0.73$, $q = 0.175$
and $A=0.353$ are used in the simulations, following \cite{Jenkins2001MNRAS.321..372J}. 
The critical overdensity in the definition of $\nu$ above 
involves Sheth-Tormen correction following \cite{Sheth:2001dp}. 
Note that the HMF given by eq.~\eqref{eq:dndM} varies with redshift with the redshift dependence entering via the terms that depend on $D(z)$ on the right hand side. Overall, there is an amplitude increase as the redshift decreases which is proportional to $D^{-2}(z)$, while  the shape also gets scaled via the $D$ factors that enter in eq.~\eqref{eq:ST_term}. 

\begin{figure}[tbp]
    \centering
    \begin{subfigure}{0.48\textwidth}
        \centering
        (a)\\
        \includegraphics[width=0.9\textwidth, height=5 cm]{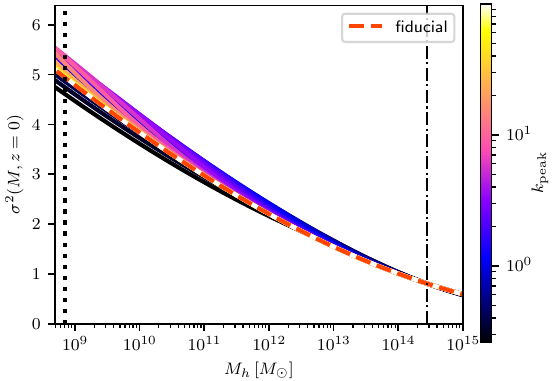}\\
        \vskip\baselineskip
        (b) \\
        \hskip -1.8cm
        \includegraphics[width=0.94\textwidth, height=13cm]{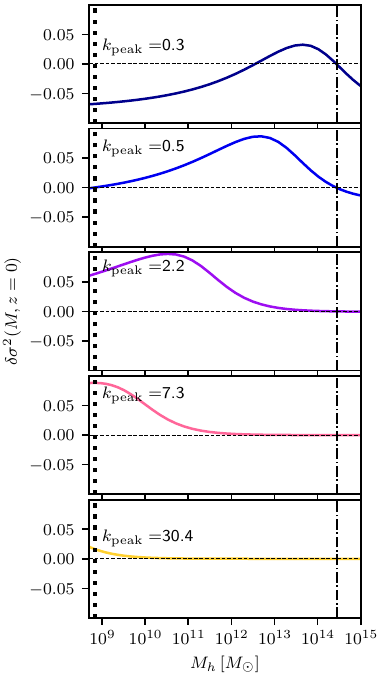}
    \end{subfigure}
    \hfill
    \begin{subfigure}{0.48\textwidth}
        \centering
        (c) \\        
        \includegraphics[width=1.05\textwidth, height=5cm]{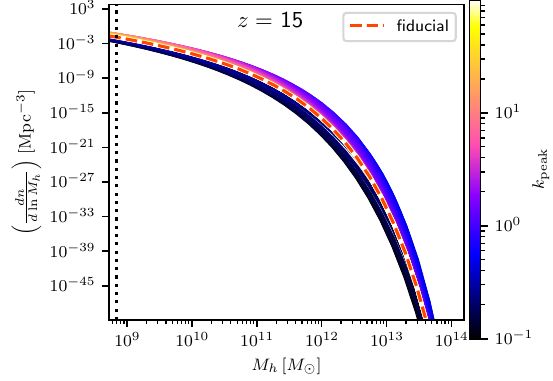}\\
        \vskip\baselineskip
        (d)\\
        \hskip -2mm
        \includegraphics[width=0.88\textwidth, height=13cm]{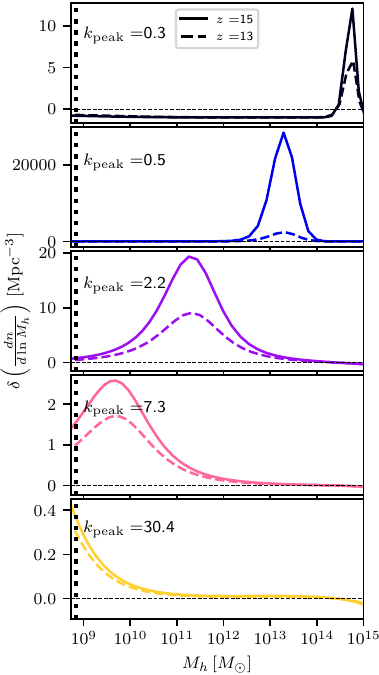}
    \end{subfigure}
     \caption{(a) $\sigma^2(M)$ extrapolated to $z=0$ for the fiducial and bump models corresponding to our range of $k_{\rm peak}$. The black dotted vertical line indicates  $M_{\rm min}=5\times 10^{8} M_{\odot}$, while the dot-dashed lines indicate the halo mass value $2.8\times 10^{14} M_{\odot}$  corresponding to $\sigma_8$. 
     (b) The fractional difference $\delta\sigma^2(M)$ between bump and fiducial models.
     (c) The HMFs for the fiducial 
    and bump models are shown for $z \sim 15$. 
    The bump-like features have amplitude $A_{\rm I} = 10^{-9}$. The fiducial model is shown in orange color. The black dotted vertical line indicates  $M_{\rm min}$. 
    (d) The fractional differences of the HMF curves 
    for the bump models in panel (b) 
    relative to the fiducial model are shown for two redshifts, $z=15$ and 13. 
    The turnover scale in this case is $k_{\rm peak} = 0.5\, {\rm Mpc}^{-1}$.}
    \label{fig:ST_HMF_bumps}
\end{figure}

{Eq.~\eqref{eq:dndM} tells us that the effect of the bump model on the HMF can be traced back to $\sigma^2(M)$. 
Before proceeding, we note that two halo-mass scales that are important for the discussion here. The first is $M_{\rm min}$, as 
given in eq.~\eqref{eq:Mmin}. 
Structure formation is driven by halos whose mass $M_h$ is above $M_{\rm min}$. Of such halos, the lower mass ones in the vicinity of  $M_{\rm min}$ are much more numerous than the high mass ones (see the plot of HMF in panel (c) of figure \ref{fig:ST_HMF_bumps}). Hence, any physical process that impacts such lower mass halos more than high mass ones will leave a relatively stronger imprint on the HMF.   
Secondly, we can associate a halo mass scale, $M_{\rm peak}^R$ with a given value of $k_{\rm peak}$ by equating the right-hand side of eq.~\eqref{eq:tophat} to $2\pi/k_{\rm peak}$. {Thus, $k_{\rm peak}$ and $M_{\rm peak}^R$ are anti-correlated (see figure~\ref{fig:k_M} in appendix~\ref{app:kpeak_m}).} We can anticipate that the effect of the primordial bump on physical observables  will typically peak around  this mass scale. The peak will be accompanied by a corresponding range of $M_h$ over which the effect will be felt\footnote{As the width of the bump-like features in $k-$space is directly proportional to its location $k_{\rm peak}$,  a range of mass scales will be 
influenced on the HMF.}.  The location of $M_{\rm peak}$ and the range of influence relative to $M_{\rm min}$ will determine the effect the primordial bump has on the HMF.}
{We now quantify how the bump models affect the HMF,  which in turn will explain the effect on $\left<x_{\rm HI}\right>$ and $\left<\delta T_b\right>$ observed in the previous subsection.} 

%

{In panel (a) of figure~\ref{fig:ST_HMF_bumps},  we plot $\sigma^2(M)$ 
 obtained from \texttt{21cmFAST} for the fiducial 
and bump models extrapolated to $z=0$. $M_{\rm min} = 5\times 10^8\,M_{\odot}$ is shown by the black vertical dotted line. The black dot-dash line marks the mass value $M \simeq 2.8 \times 10^{14}\, M_\odot$, which corresponds to the 
scale of the power spectrum normalization, $8\,h^{-1}{\rm Mpc}$.
The values of $\sigma^2(M)$ for all the models coincide at this mass value. Panel (b) shows the fractional difference $\delta\sigma^2(M)$ of the bump models with respect to  the fiducial one for some values of $k_{\rm peak}$. Here, two points are worth mentioning. 
First of all, with increasing $k_{\rm peak}$, 
the peak of the profiles, $M_{\rm peak}^{\sigma^2}$, shifts toward the low-mass side, 
which can be understood by the relation between 
$k_{\rm peak}$ and $M_{\rm peak}^R$, as discussed earlier (see appendix~\ref{app:kpeak_m} for details). 
Secondly, there is an overall shift up of the entire profile on the  vertical axis, 
which is primarily due to the normalization of the power spectrum 
at $8\,h^{-1}{\rm Mpc}$.\footnote{%
{
The resulting shift in 
the amplitude of the nearly scale-invariant part of the power spectrum,
$A_s$ in eq.~\eqref{eq:PS_bump_ch4}, 
will be constrained by the CMB observations.}
}
Furthermore, $k_{\rm peak} = 0.5 = k^{\rm turn}$ roughly marks the value where the left tail of the profile transitions from negative to positive values 
of $\delta \sigma^2(M, z=0)$ at $M_h=M_{\rm min}$. } 

{Next, we examine the effects of primordial bump-like features on the  HMFs. Panel (c) of figure~\ref{fig:ST_HMF_bumps} shows the HMF curves for the fiducial 
and bump models for the same range of $k_{\rm peak}$ as the left column. $M_{\rm min}$ is again shown by the black vertical dotted lines.
Panel (d) shows the fractional differences of the HMF of the bump models relative to the fiducial model  
for various values of $k_{\rm peak}$, for $z=15$ and 13.} 
%
%
{With increasing $k_{\rm peak}$, 
the peak of the profile, $M_{\rm peak}^{\rm HMF}$, shifts toward the low-mass scale, accompanied by a broadening of the profile.  The peak locations $M_{\rm peak}^{\rm HMF}$ are roughly an order of magnitude higher than $M_{\rm peak}^{\sigma^2}$ 
(refer to figure~\ref{fig:k_M} in appendix~\ref{app:kpeak_m}). This is because the HMF is related to the derivative of $\sigma(M)$. The broadening of the profiles with increasing $k_{\rm peak}$  is due to the factor $\nu f(\nu)$ in the expression for the HMF.   Again, $k_{\rm peak} = 0.5 = k^{\rm turn}$ roughly marks the value where the left tail of the profile transitions from negative to positive values. We find a drop in the amplitude from $z=15$ to 13 without changes in the peak locations. } 
{The implication of the above-observed behaviour of the HMFs is that, for $k_{\rm peak} < k^{\rm turn}$, there will be fewer halos having $M_h > M_{\rm min}$ for the bump model relative to the fiducial power spectrum. This means that, at a given $z$,  the bump model will be in a less advanced stage of reionization compared to the fiducial model.  For $k_{\rm peak} > k^{\rm turn}$, the situation is reversed, and we should obtain a more advanced reionization for the bump model. This trend is exactly what we observe in figure \ref{fig:global21_all}.}
The effects of primordial features on the collapsed fraction of halos also show a similar trend
and is discussed in appendix~\ref{subsec:fcoll}.

To summarize, the effect of the primordial bump-like features on  
observable quantities such as ionization history and  
global 21 cm profile is driven by their effects on the HMF
(primarily on the low-mass scales set by $M_{\rm min}$), which 
further depends on  $\sigma^2(M)$. 
Due to the power spectrum normalization at 8 $h^{-1}{\rm Mpc}$, 
the enhancement of power at different scales introduced by the 
primordial features significantly alters the number density of halos and, consequently, the structure formation.  
We find a particular scale where the effects of primordial features 
do not show up in the global 21 cm profile, i.e., the turnover scale, $k^{\rm turn}$. 
The turnover scale is also crucial as the bump-like features 
with $k_{\rm peak} > k^{\rm turn}$ induce early reionization, 
and $k_{\rm peak} < k^{\rm turn}$ result in late reionization. 
Since the value of $M_{\rm min}$ is influenced by 
the underlying EoR model chosen through the 
virial temperature $T_{\rm vir}$, 
it is expected that the value of $k^{\rm turn}$ 
may depend on $T_{\rm vir}$, which we explore in the next section.

\section{Global 21 cm signal as a probe of bump-like 
primordial features with varying astrophysical parameters}
\label{sec:s5}

{Having understood the effect of bump-like primordial features 
on the global 21 cm signal and the existence and physical origin 
of $k^{\rm turn}$ for fixed astrophysical parameters, we now turn our attention 
to investigating the effects when the astrophysical parameters are also varied. 
This exercise will reveal {how $k^{\rm turn}$ will depend on the EoR parameters and} any degeneracies and distinguishability of 
the effects of primordial features from the effects of different astrophysical parameters. }
We focus on three astrophysical parameters that crucially impact  
the ionization history and global 21 cm profile:
the ionizing efficiency $\zeta$, the virial temperature of the halos $T_{\rm vir}$, 
and the soft-band X-ray luminosity per unit SFR $L_X$. 

 \subsection{Dependence of the turnover scale 
 $k^{\rm turn}$ on the choice of astrophysical parameters}
 \label{subsubsec:kinv}

\begin{figure}[tbp]
    \centering
    \includegraphics[width=0.48\textwidth]{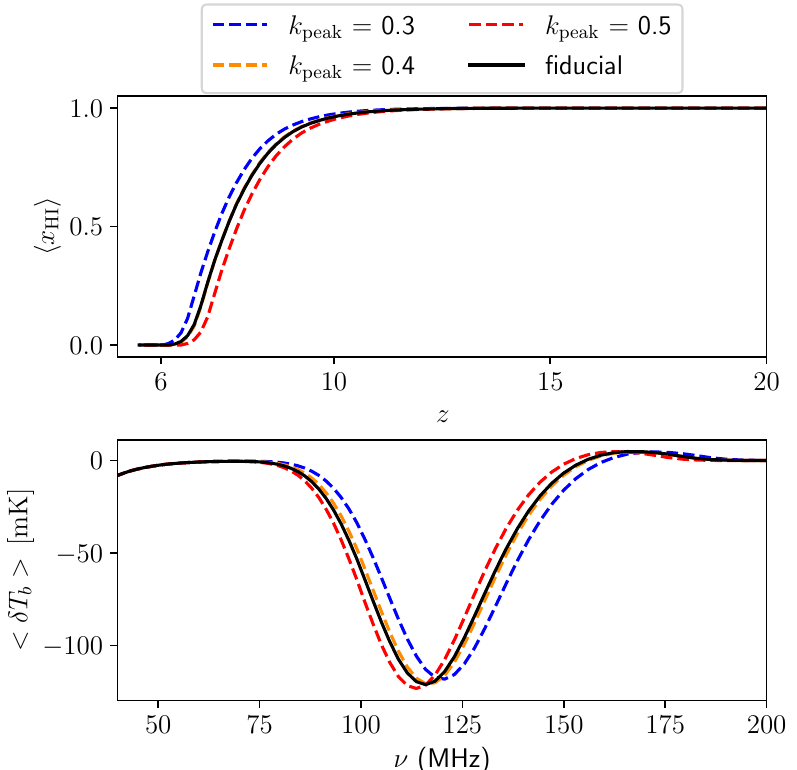}
      \includegraphics[width=0.45\textwidth]{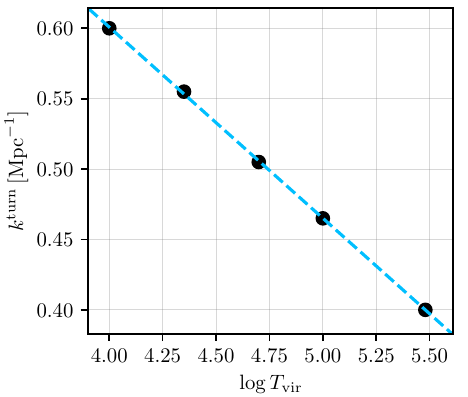}
\caption{[Left] The ionization history and the global 21 cm profile 
for the bump models using the ``bright'' galaxies model 
as the fiducial. The amplitude of the features is fixed at 
$A_{\rm I} = 10^{-9}$. 
[Right] Simulated values of $k^{\rm turn}$ for different  $T_{\rm vir}$ values.
$L_X$ is fixed to $40.5$ for this plot. 
The cyan dashed line represents a linear fit.}
    \label{fig:global_BrightGal}
\end{figure}
%
%
%

{For varying the EoR parameters, we follow the convention of ``faint'' and ``bright'' galaxies models, 
as given in \cite{Greig:2017jdj}. In the ``faint'' galaxies model, 
the EoR parameter values are roughly $\zeta \sim 30$ and $T_{\rm vir} \sim 5 \times 10^4\, {\rm K}$. 
The fiducial EoR model used in the previous section falls in this category. 
In contrast, in the ``bright'' galaxies model, 
structure formation is primarily driven 
by rare and relatively massive galaxies
with the EoR parameter values $\zeta \sim 200$ and $T_{\rm vir} \sim 3 \times 10^5\, {\rm K}$. In the latter case, we have $M_{\rm min} \approx 8 \times 10^9\, M_\odot$ (roughly an order of magnitude higher than the faint galaxies model).} 
%
{Consequently, for such models, the influence on structure formation can be deciphered from the behaviour of the HMF 
at mass scales $M_h \gtrsim 8 \times 10^9\, M_\odot$. }

{Since $k_{\rm peak}$ and $M_h$ are inversely related,  
an increase in $M_{\rm min}$ is expected to  
cause a decrease in $k^{\rm turn}$.  
To verify this, we calculate the ionization history 
and global 21 cm profile for different values of $T_{\rm vir}$. 
{Note that since $T_{\rm vir}$ defines the mass range of galaxies 
involved in structure formation, in each case, the ionizing efficiency 
parameter $\zeta$ is appropriately adjusted 
to ensure that the optical depth to reionization, $\tau_e$,  
{is within} the 95\% limits measured by CMB observations 
from {\it Planck}. We fix $L_X$ to be $40.5$}. 
The turnover scale $k^{\rm turn}$ is chosen as the value of $k_{\rm peak}$
for which the redshift of the minimum
in the absorption trough of its 
global profile matches that of the fiducial model.
The results are plotted in figure~\ref{fig:global_BrightGal}.  On the left column, we plot redshift evolution of $\langle x_{\rm HI}\rangle$ and $\langle \delta T_b \rangle$ for the bright galaxies model with parameter values $\zeta = 200$ and $T_{\rm vir} = 3 \times 10^5\, {\rm K}$ having different values of $k_{\rm peak}$.  We observe in this case that the turnover scale is now 
$0.4\, {\rm Mpc}^{-1}$, which, as expected, is lower than the value for the faint galaxies model. }

{The right column of figure~\ref{fig:global_BrightGal}   shows $k^{\rm turn}$ as a function of $T_{\rm vir}$.  } 
{The black dots are the values of $k^{\rm turn}$ derived 
from the simulations. From left to right, $\zeta$ is increasing from 10 to 200.} The blue dashed line represents a
straight-line fit\footnote{The best-fit parameters are 
$\alpha = 1.144$ 
and $\beta = -0.136$. 
$\alpha$ and $\beta$ are likely to be physical quantities, and we will explore their meaning and the linear relation in the future.} given by the form $k^{\rm turn} = \alpha + \beta(\log T_{\rm vir})$.

We repeated the calculations for various values of $L_X$ and found that its impact on $k^{\rm turn}$ is negligible. This is because $k^{\rm turn}$ is affected only when changes in astrophysical parameters cause a shift in the global profile compared to the fiducial case. While $L_X$ influences heating and alters the depth of the absorption trough, $T_{\rm vir}$ causes a shift in the global profile.
Thus, we conclude that for a given EoR model, 
there exists a specific scale $k^{\rm turn}$, 
majorly governed by $T_{\rm vir}$,
at which the bump-like features 
show negligible effects on 
the global 21 cm profile.



\subsection{Comparison of the effects of bump models 
with EoR models}
\label{subsec:EoRmodels}
%
In this section, we explore how changes in EoR models, 
characterized by $\zeta,\, T_{\rm vir}$ and $L_X$, 
affect the ionization history and the global 21 cm profile. 
The purpose is to compare such effects with the influence of bump-like features. 
We choose the parameter ranges for the EoR models as  
$\zeta \sim [10, 50]$, 
$\log T_{\rm vir} \sim [4.4, 5.4]$, and 
$\log L_X \sim [38, 42]$. {These ranges ensure that} the predicted ionization 
history and 21 cm signal are consistent 
with theoretical expectations and observational constraints 
\cite{Greig:2017jdj}. 

\begin{figure}[tbp]
    \centering
    (a) Ionization history \\
    \includegraphics[width=0.95\textwidth]{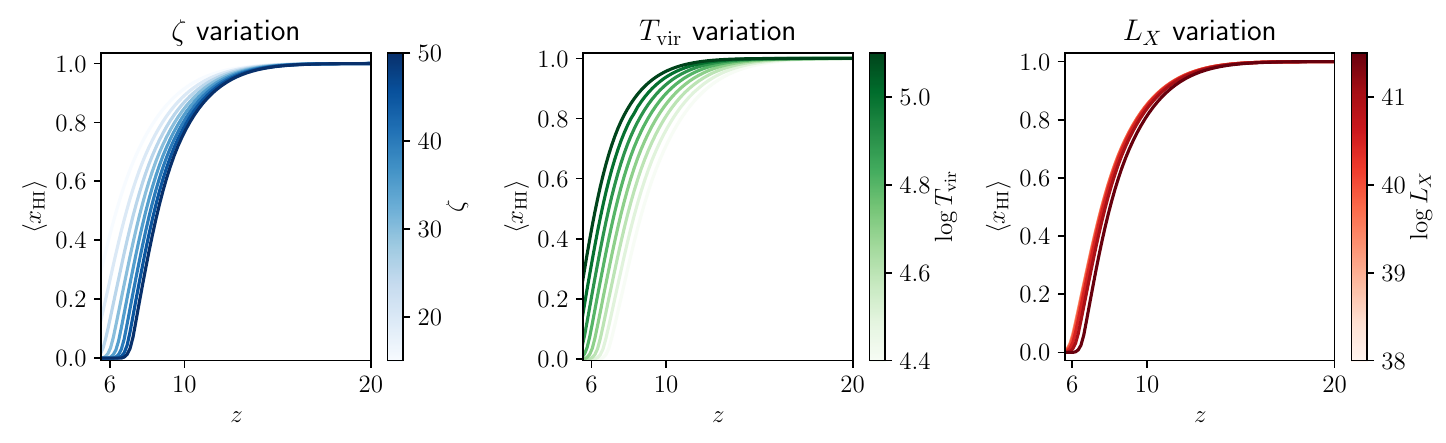}\\
    (b) Global 21 cm profile \\
    \includegraphics[width=\textwidth]{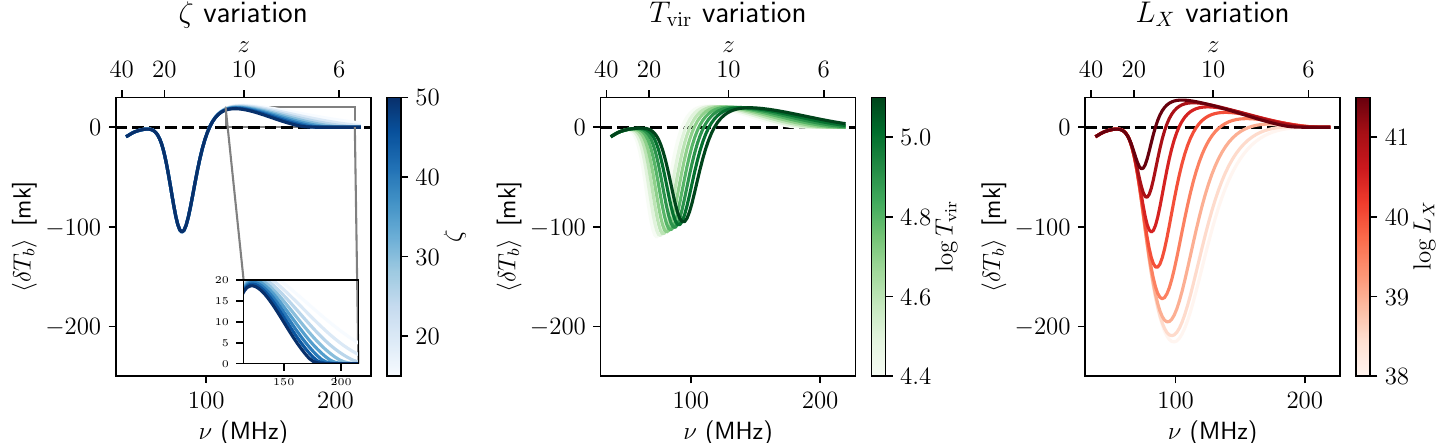}
    \caption{[Top] Ionization history and [bottom] global 21 cm profiles 
for the EoR models, illustrating the effects of varying 
$\zeta$ (left), $T_{\rm vir}$ (middle), and $L_X$ (right).
Increasing shade represents an increase in the magnitude of the parameter.}
    \label{fig:global_EoR}
\end{figure}

Figure~\ref{fig:global_EoR} shows 
the effect of varying each EoR parameter, {keeping the other two fixed at their fiducial values,} on the reionization history (top panel) 
and the global 21 cm profile (bottom panel). {Below, we further quantify the dependence of these observables on the EoR parameters.}


%
%

\subsubsection{Ionization history}
\begin{figure}[tbp]
    \centering
    \includegraphics[width=1\textwidth]
    {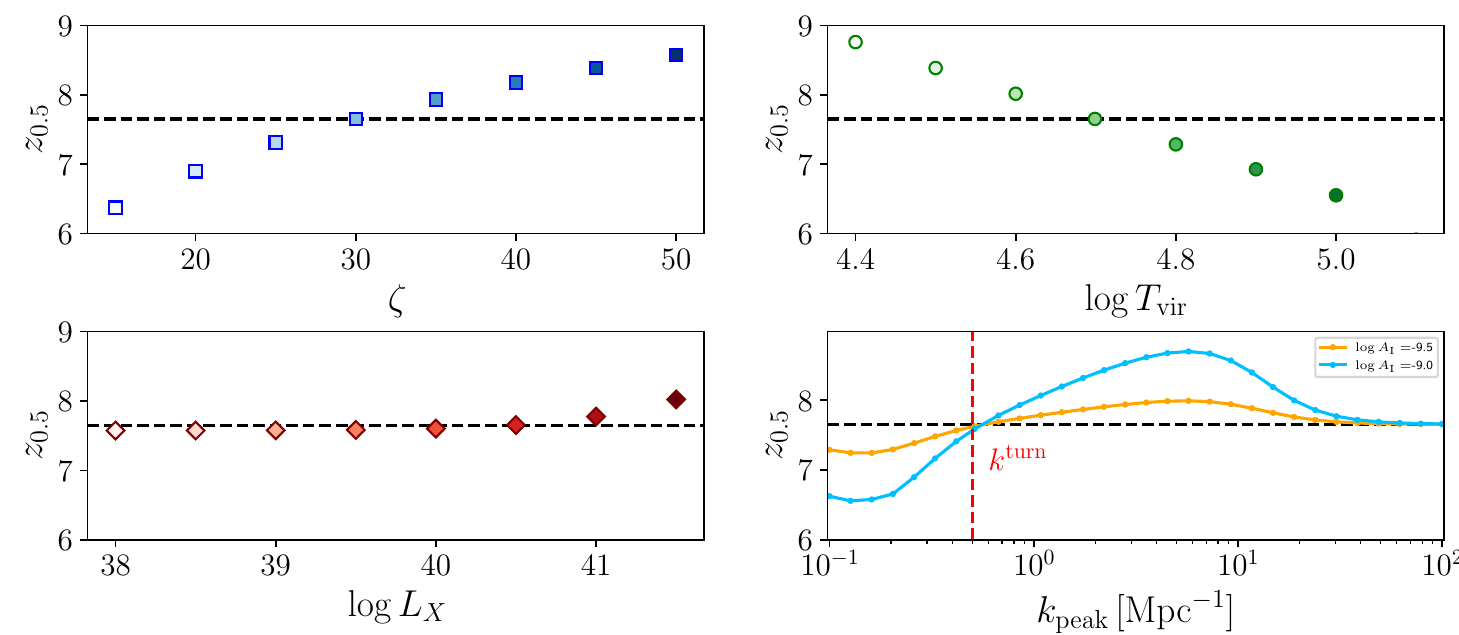}
\caption{The redshifts corresponding to 50\% reionization, $z_{0.5}$, 
plotted against the parameters of EoR and bump models.
The horizontal dashed line represents $z_{0.5}$ for the fiducial model. }
    \label{fig:z50}
\end{figure}

We begin by quantifying the ionization history by
plotting the redshift corresponding to 50\% ionization, $z_{0.5}$ 
(i.e., $\left<x_{\rm HI}\right> = 0.5$) 
against different parameters, as shown  {in the top and bottom left panels of}  figure~\ref{fig:z50}.
Increasing $\zeta$ 
results in earlier reionization, leading to a higher $z_{0.5}$. 
In contrast, 
increasing $T_{\rm vir}$ results in a lower $z_{0.5}$. 
Since higher $T_{\rm vir}$ corresponds to larger $M_{\rm min}$ (eq.~\eqref{eq:Mmin}), 
 structure formation is driven by relatively massive galaxies 
with lower number density, delaying the end of reionization. 
We also find that $L_X$ has little influence on the reionization history. 
{For comparison, the effect of $k_{\rm peak}$ for bump-like features 
is shown in the bottom right panel for two values of $A_I$. We observe a trend distinct from the trend seen for the EoR parameters.} 
With an increase in $k_{\rm peak}$, 
$z_{0.5}$ increases and reaches the value of the fiducial model 
at $k \sim k^{\rm turn}$.  
{Beyond that, $z_{0.5}$ further increases} 
till it reaches a maximum when the mass scale 
corresponding to $k_{\rm peak}$, i.e., $M_{\rm peak}$, 
coincides with $M_{\rm min}$ and results in 
a maximum number density of low-mass halos in the HMF 
(refer to figure~\ref{fig:ST_HMF_bumps} (d)).
As $k_{\rm peak}$ increases beyond this scale, 
only the tail of the bump affects the HMF, and consequently, 
$z_{0.5}$ decreases. 
Increasing the amplitude of the bump feature enhances the effect of 
$k_{\rm peak}$. 
{Visually,} we notice that for low values of $k_{\rm peak}$, 
the parameter $\zeta$ shows correlation, and $T_{\rm vir}$
shows anti-correlation with $k_{\rm peak}$. 
%

\subsubsection{Global 21 cm profile}
We parameterize the global 21 cm profile in two ways:
\begin{enumerate}
    \item[(a)] Fitting the absorption trough 
    of the signal to a Gaussian profile, characterized by 
    amplitude $T_{b_{\rm min}}$, 
    mean frequency $\nu_{\rm min}$, and 
    standard deviation $\sigma$. 
    \item[(b)] Identifying the average brightness temperatures $\left<\delta T_b\right>$ 
    and corresponding frequencies $\nu$ (or redshifts) at 
    three key extrema points
    following \cite{Cohen:2016jbh}.
\end{enumerate}
{The findings are discussed below.}

\vskip .3cm
\noindent{\bf (a) Fitting a Gaussian profile:} 
We fit the absorption trough of the global profile to a  
Gaussian function and obtain 
$T_{b_{\rm min}}$, 
$\nu_{\rm min}$ and $\sigma$. 
\begin{figure}[tbp]
    \centering
    \includegraphics[width=\textwidth]
    {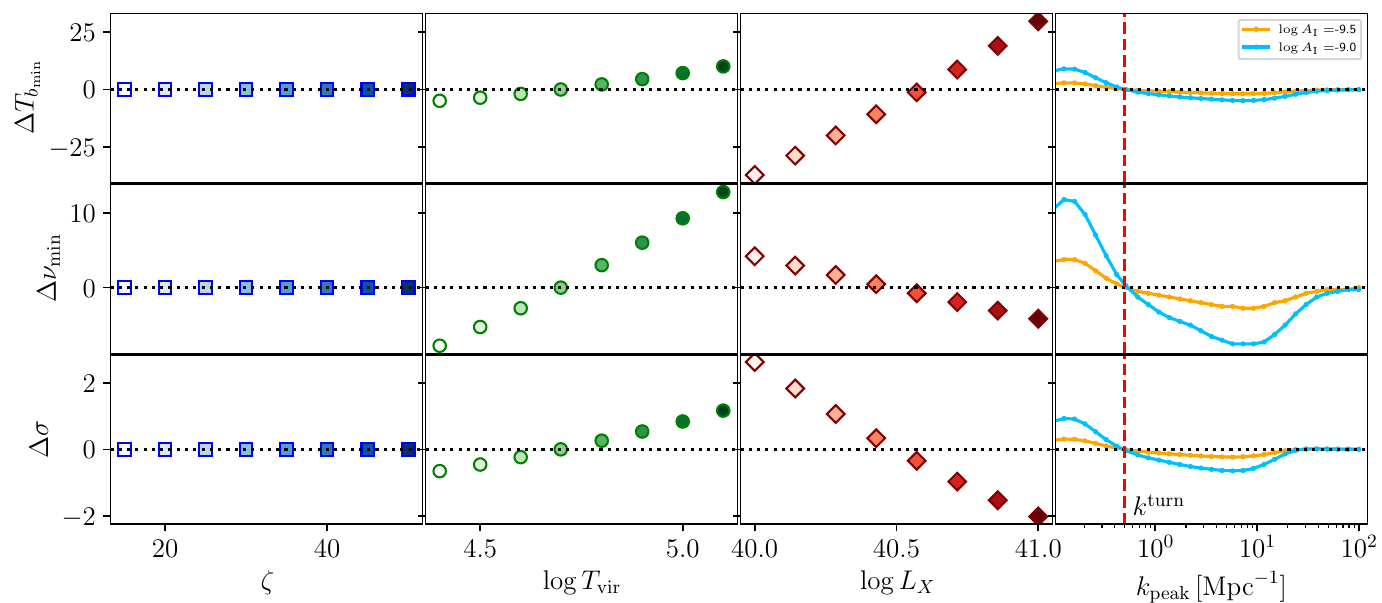}
    \caption{The relative shift in the parameters - depth of the absorption profile, 
$\Delta T_{b_{\rm min}}$ (top), frequency of the absorption 
trough, $\Delta \nu_{\rm min}$ (middle), and width of 
the profile, $\Delta \sigma$ (bottom), 
relative to the fiducial model, obtained by fitting a 
Gaussian profile.}
    \label{fig:fwhm_EoR_bumps}
\end{figure}
The shift in these quantities for the models 
with varying EoR parameters or bump parameters 
relative to the fiducial model, i.e., 
$\Delta X =  X^{\rm EoR/bump} -  X^{\rm fid}$, where 
$X$ is $T_{b_{\rm min}}$, 
$\nu_{\rm min}$ and $\sigma$,
are plotted in the three 
rows of figure~\ref{fig:fwhm_EoR_bumps}, 
respectively. 

As {seen} in the bottom panels of figure~\ref{fig:global_EoR}, 
$\zeta$ primarily affects the EoR at $z < 15$, 
having minimal influence on the parameters defining 
the absorption trough. Increasing $T_{\rm vir}$ 
results in a profile shift toward higher frequencies 
(lower $z$), causing the absorption trough to 
become wider and shorter. 
The parameter $L_X$ 
significantly {alters the depth and width} of the global 21 cm profile. 
As X-ray luminosity heats up the IGM, 
a higher value of $L_X$ rapidly raises the kinetic temperature, 
which is coupled to the spin temperature 
$T_S$, above the background temperature $T_\gamma$, 
leading to an earlier emission signal. 
Thus, an increase in $L_X$ contributes to 
the narrowing of the width and reduces 
the depth of the profile. {These effects of $\zeta,\, T_{\rm vir}$ and $L_X$ are quantified and illustrated in 
the first three columns of figure~\ref{fig:fwhm_EoR_bumps} (left to right).}

{For comparison with the above,} the effects of bump-like features are shown 
in the last column of figure~\ref{fig:fwhm_EoR_bumps} {for the same two values of $A_{\rm I}$ as in figure \ref{fig:z50}.} 
As previously noted,  
if $k_{\rm peak} < k^{\rm turn}$, 
the number density of halos is lower than that of 
the fiducial model, leading to delays in heating 
and reionization as described in section~\ref{subsubsec:HMF}. 
Consequently, the global 21 cm profile shifts toward 
the high-frequency or low-$z$ end. 
In contrast, if $k_{\rm peak} > k^{\rm turn}$, 
the number density of halos exceeds that of the 
fiducial model, 
resulting in a shift of profile toward 
the low-frequency. These effects become 
more pronounced with an increase in the amplitude of the 
primordial features. 
%

{As stated earlier, the above exercise aims to determine whether the  
primordial bump-like features can be distinguished from EoR parameters.  First,} we note that the correlation of $\zeta$ with $k_{\rm peak}$
noticed in the ionization history in figure~\ref{fig:z50} does not show up in 
the global profile.  
{This indicates that the effects of primordial features can potentially be 
distinguished from the effects of $\zeta$. 
Secondly, from a comparison between columns 3 and 4 of 
figure~\ref{fig:fwhm_EoR_bumps}, we find that the effect of $L_X$ can be clearly distinguished 
from $k_{\rm peak}$. 
Finally,} the anti-correlation of 
 $T_{\rm vir}$ with $k_{\rm peak}$ 
 noticed in figure~\ref{fig:z50} for the low $k_{\rm  peak}$ values
 remains in figure~\ref{fig:fwhm_EoR_bumps}. 
This means that
changes to the global profile, such as 
increasing the depth of the absorption 
tough, shifting $\nu_{\rm min}$ to higher frequency and 
broadening the profile,  
all of these can be achieved either by 
increasing $T_{\rm vir}$ or by introducing a primordial bump-like feature
with a smaller $k_{\rm peak}$. 

\vspace{2mm}
\noindent{\bf (b) Points of extrema:} 
\begin{figure}[tbp]
    \centering
    \includegraphics[width=0.8\textwidth]{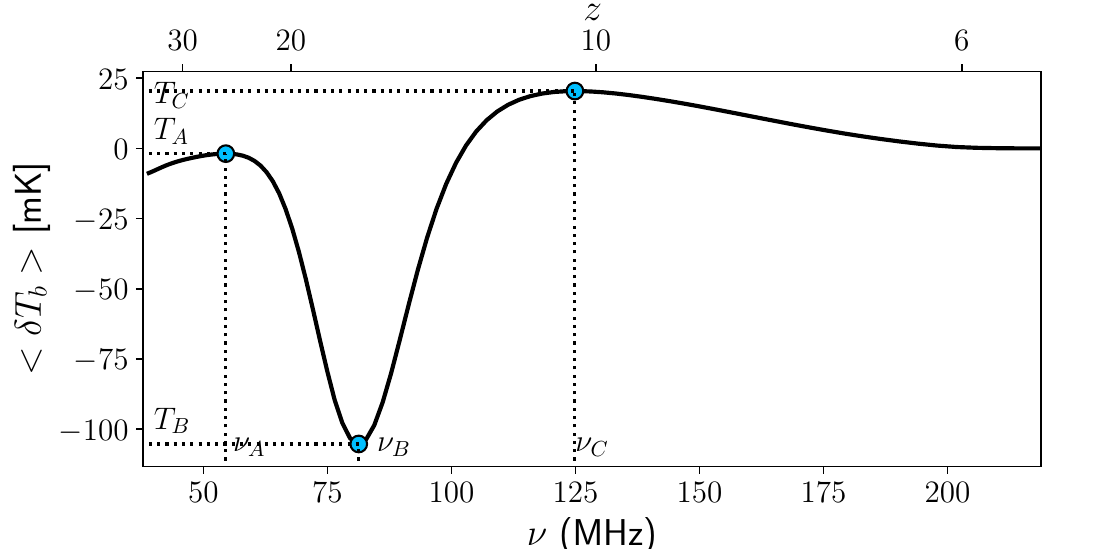}
    \caption{The points of extrema used for quantifying the 
    global 21 cm signal in figure~\ref{fig:T_v_nu}.}
    \label{fig:fid_abc}
\end{figure}
\begin{figure}[tbp]
    \centering
    (a) EoR models\\
    \includegraphics[width=\textwidth]
    {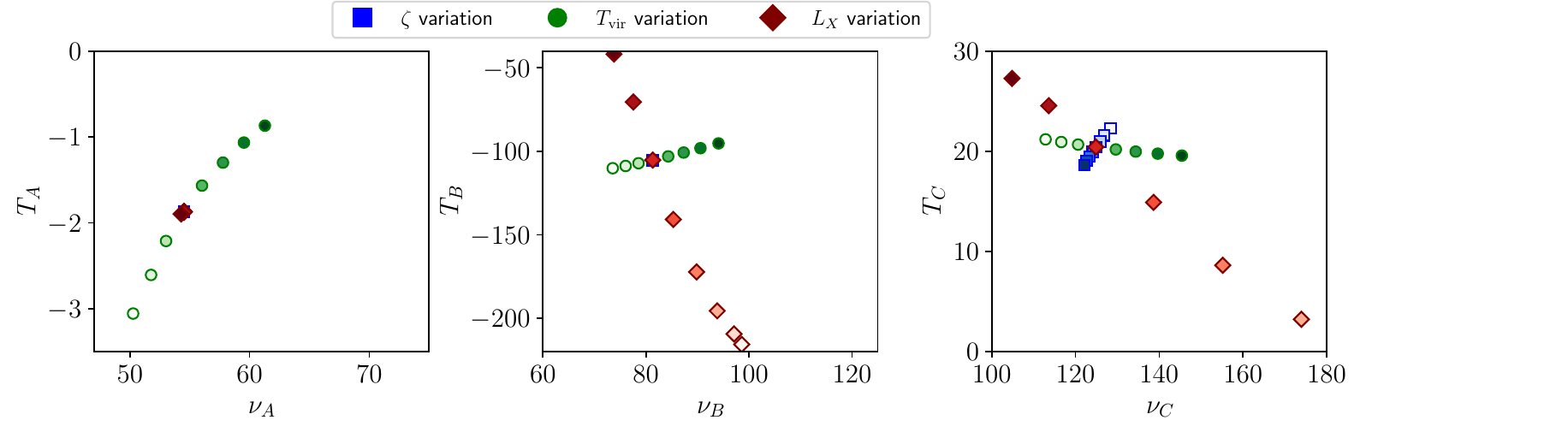}\\
    (b) Bump models\\
    \includegraphics[width=1\textwidth]
    {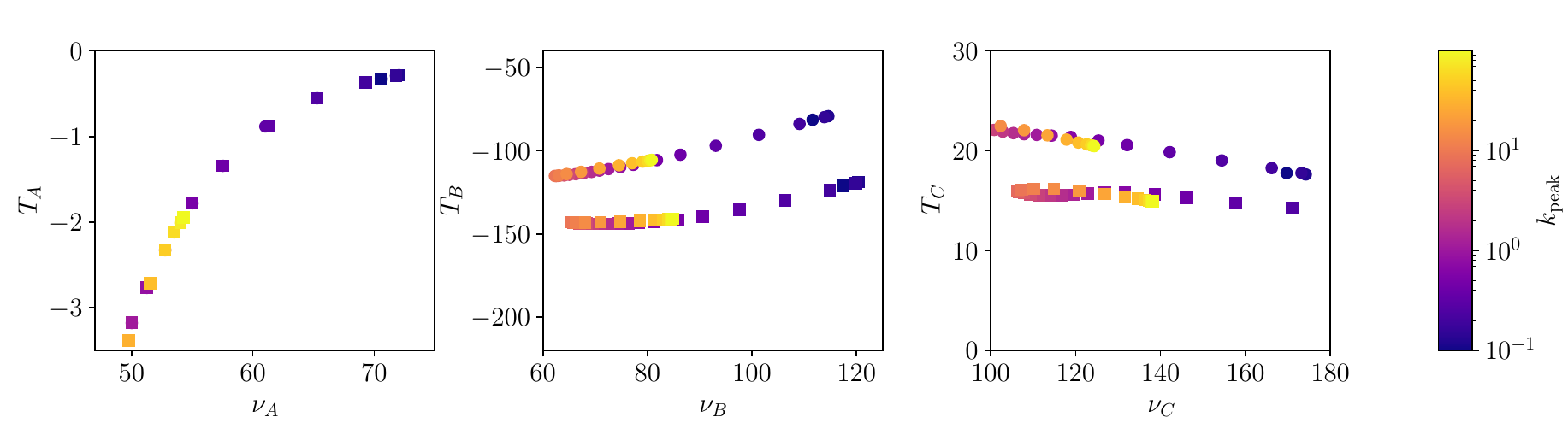}
    \caption{The average brightness temperature and frequency corresponding 
    to points A, B and C in figure~\ref{fig:fid_abc} are plotted in 
    the left, middle and right panels, respectively. 
    Plots in (a) show the variation of astrophysical parameters $\zeta$, $\log T_{\rm vir}$, $\log L_X$ and plots in (b)
    show the variation in $k_{\rm peak}$ with $A_{\rm I} = 10^{-9}$. 
    Increasing shade represents increase in magnitude of the parameter.
    The bump models are plotted for two sets of fiducial values: 
    $L_X = 40.5$ is denoted by circles and 
    $L_X = 40.0$ is denoted by squares.}
    \label{fig:T_v_nu}
\end{figure}
We now quantify the global 21 cm profile 
by noting the average brightness temperatures 
and frequencies corresponding to the points 
of two maxima and one minimum
following \cite{Cohen:2016jbh}. 
We refer to these points as {$T_A$ at $\nu=\nu_A$, 
$T_B$ at $\nu=\nu_B$, and $T_C$ at $\nu=\nu_C$}, 
as shown in figure~\ref{fig:fid_abc}.

The top panel of figure~\ref{fig:T_v_nu} shows 
temperatures 
and frequencies related to points A, B, and C
for the 
EoR models, while the 
bottom panel shows the same for bump models. 
The effect of $\zeta$ is significant only 
at point `C', corresponding to the maximum 
of the profile in the low-redshift regime. 
The EoR models with varying $L_X$ 
are placed along the diagonal of the plots 
for `B' and `C', showing no scatter 
in the plot of `A'. 
However, $T_{\rm vir}$ shows distinct trends 
in all three plots.
On the bottom panel, the effect of 
variation of $k_{\rm peak}$ is 
shown for two 
different underlying EoR models: $L_X = 40.5$ 
is denoted by circles and $L_X = 40.0$ 
is denoted by squares. 
The points corresponding to 
the bump-like features 
with $k_{\rm peak} > k^{\rm turn}$ 
reflect as overlapping points in all the three plots of A, B and C, 
indicating degeneracy among the bump models on these scales. 

Once again, in all of these plots, the bump models show 
a trend similar to that of $T_{\rm vir}$, 
while the effects of $\zeta$ and 
$L_X$ are clearly distinguishable.
The degeneracy of the bump model with 
the virial temperature was expected as 
$T_{\rm vir}$ directly influences 
the effects of bump-like features 
on the global 21 cm profile
through the mass threshold $M_{\rm min}$.
Therefore, higher-order statistics, such as the power 
spectrum may help break the degeneracy, 
which will be discussed in an upcoming work. 
%

\section{Constraining bump-like primordial features  
{from Planck constraints on reionization history}} 
\label{subsec:constraints}
In the previous sections, we emphasized that 
primordial bump-like features 
in the range 
$10^{-1} < k_{\rm peak}\, [{\rm Mpc}^{-1}] < 10^2$
can significantly affect the ionization history and the 
global 21 cm profile. 
In this section, we explore constraints on the 
parameters $A_{\rm I}$ and $k_b$
by comparing the simulations with  
the currently available observations
on the reionization history. 
All the astrophysical parameters are fixed to their fiducial values in this analysis. 

The CMB observations by {\it Planck} \cite{Planck:2015sxf, Planck:2018vyg} 
have imposed limits on the ionization of the high-redshift 
intergalactic medium (IGM) through the parameter - Thomson scattering optical depth,
$\tau_e$. 
As ionization progresses following the formation of the first stars and galaxies, 
CMB photons are scattered by free electrons in the IGM. 
Consequently, the observed CMB is influenced by the total column density of these 
free electrons along the line of sight. 
The reionization optical depth $\tau_e$ is
constrained as one of the six parameters of the $\Lambda$CDM model. 
CMB observations by Planck indicate $\tau_{\rm Planck} = 0.058 \pm 0.012$ (95\% CL) 
\cite{Planck:2018vyg}. 
We compute the values of $\tau_e$ for bump-like models using \texttt{21cmFAST}
and compare with $\tau_{\rm Planck}$ to derive constraints on the parameters $A_{\rm I}$ and $k_b$. 

\begin{figure}[tbp]
    \centering
    \includegraphics[width=0.65\textwidth]{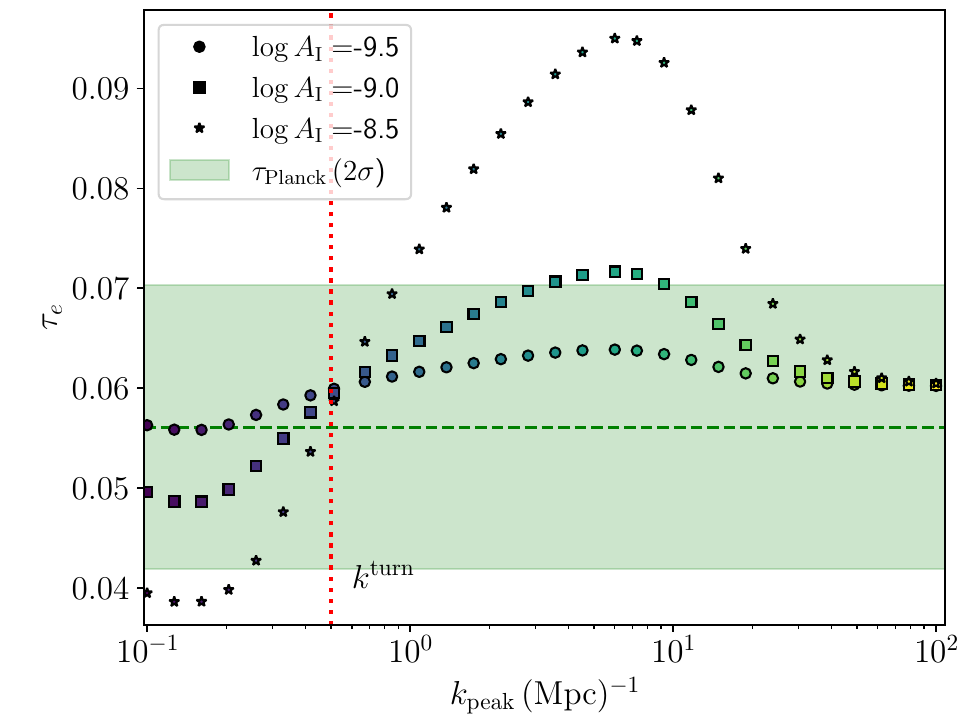}
    \includegraphics[width=0.65\textwidth]{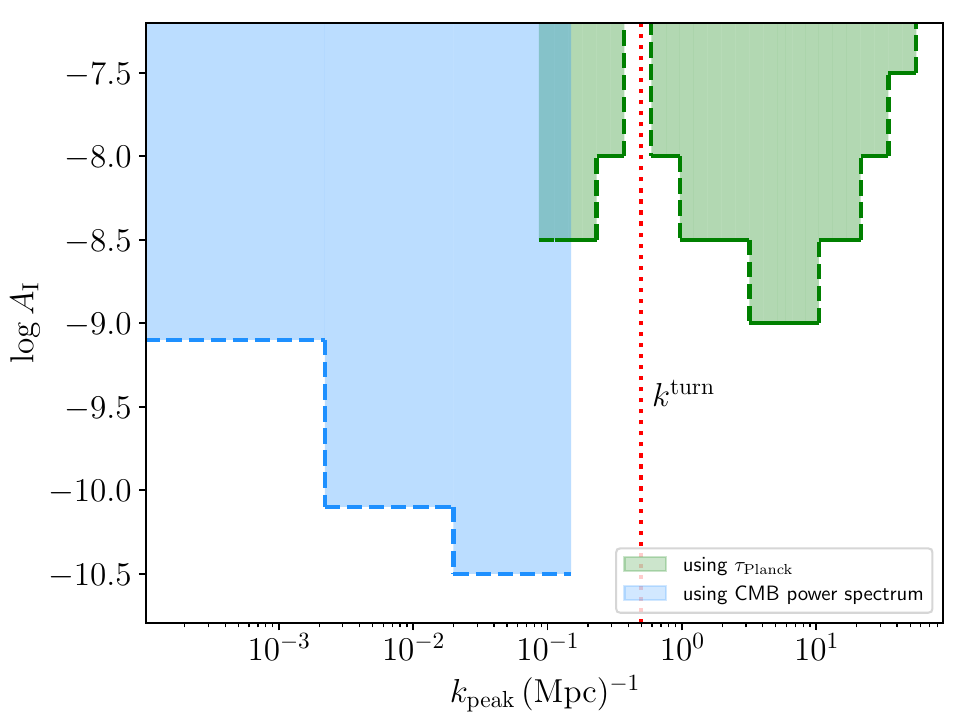}
\caption{[Top] The optical depth to reionization, $\tau_e$, predicted for bump-like models, with astrophysical parameters fixed to their fiducial values. The constraint on $\tau_e$ from Planck is shown by the dashed line, with the 95\% confidence interval indicated by the shaded region.
[Bottom] The upper limits on $A_{\rm I}$ in the range 
$10^{-4} < k\,[{\rm Mpc}^{-1}] < 0.15$ obtained from CMB data analysis 
in ref.~\cite{Naik:2022mxn} are shown by the blue shaded region.
The new constraints on $A_{\rm I}$ in the range $10^{-1} < k\,[{\rm Mpc}^{-1}] < 10^2$
by comparing the $\tau_e$ values in this work are shown by the green shaded 
region.
{Note that the upper limits on primordial features obtained in this manner 
depend on the underlying astrophysical model, and the plotted constraints 
are valid only for the fiducial astrophysical parameter values.}
}
    \label{fig:tau_values}
\end{figure}
Our results are shown on the top panel of Figure \ref{fig:tau_values} 
with {\it Planck} constraints on $\tau_e$ shown by the 
shaded region. 
Different markers indicate the amplitude of the features
$A_{\rm I}$ at a given $k_{\rm peak}$. 
Within the 95\% confidence interval, 
several bump-like models predict $\tau_e$ values inconsistent 
with the data, allowing us to exclude these models. 
On the bottom panel, upper limits on $A_{\rm I}$ are presented 
for each $k_{\rm peak}$ obtained from this analysis
in the range of $k\sim [10^{-1}, 10^2]$.
We also show our constraints from a detailed analysis using the CMB 
angular power spectrum (temperature and polarization data) \cite{Naik:2022mxn}
in the range $k\sim [10^{-4}, 0.15]$. 
Due to the distinct behaviour of bump-like features 
above and below $k^{\rm turn}$, 
the constraints become weaker as 
$k_{\rm peak} $ approaches $k^{\rm turn}$ and 
become unconstrained at $k \sim k^{\rm turn}$. 
The upper limit on $A_{\rm I}$ is the tightest 
when the features contribute maximally to the HMF, 
i.e., around $k \sim 7 \, {\rm Mpc}^{-1}$, 
and the amplitudes remain largely unconstrained at 
$k \gtrsim 70\,{\rm Mpc}^{-1}$. 
{We note that the observational constraints on $A_{\rm I}$ and $k_{\rm peak}$ 
depend on the underlying EoR model, as $k^{\rm turn}$ is sensitive to the virial temperature. 
Consequently, the upper limits on $A_{\rm I}$ presented here are valid only for the 
fiducial astrophysical parameter values. A more comprehensive analysis, 
involving a full exploration of the astrophysical parameter space using methods such as MCMC sampling, 
is planned for future work.}

\begin{figure}[tbp]
    \centering
    \includegraphics[width=\textwidth]{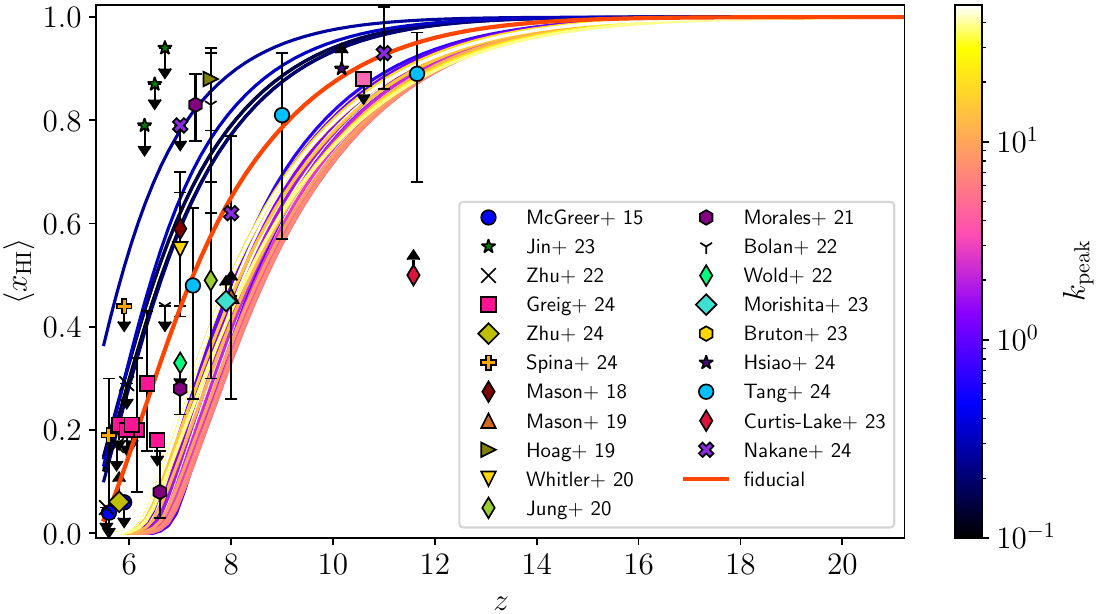}
\caption{
The solid curves show the reionization history for bump
models with different $k_{\rm peak}$ (as indicated by the color bar) 
and upper limits on $A_{\rm I}$, 
based on constraints using $\tau_e$.
The markers with error bars denote the 
95\% limits on the neutral fraction of hydrogen 
(with arrows representing upper/lower limits)
from high-$z$ quasars and Ly-$\alpha$ observations 
compiled from the literature \cite{McGreer:2014qwa,Jin:2022ngn,Zhu:2022rkx, 
Greig:2024atz,Zhu:2024huz, Spina:2024uyc, Mason:2017eqr, 
Mason:2019ixe, Hoag:2019ApJ...878...12H, Whitler:2019nul, 
Jung:2020ApJ...904..144J, Morales:2021fzt, Bolan:2022MNRAS.517.3263B, 
Wold:2022ApJ...927...36W, Morishita:2023ApJ...947L..24M, 
Bruton:2023ApJ...949L..40B, Hsiao:2023nix, tang2024jwstnirspec, 
Curtis-Lake:2023NatAs...7..622C, Nakane:2024ApJ...967...28N}. }
    \label{fig:xHI_obs}
\end{figure}

In addition to {\it Planck}, recent studies of high-redshift 
quasars and Ly-$\alpha$ observations have established constraints 
on the neutral hydrogen fraction across various redshifts. 
Figure~\ref{fig:xHI_obs} shows a compilation of 
observational constraints 
on the neutral fraction of hydrogen
from high-$z$ quasars and Ly-$\alpha$ observations 
\cite{McGreer:2014qwa,Jin:2022ngn,Zhu:2022rkx, 
Greig:2024atz, Zhu:2024huz, Spina:2024uyc, Mason:2017eqr, 
Mason:2019ixe, Hoag:2019ApJ...878...12H, Whitler:2019nul, 
Jung:2020ApJ...904..144J, Morales:2021fzt, Bolan:2022MNRAS.517.3263B, 
Wold:2022ApJ...927...36W, Morishita:2023ApJ...947L..24M, 
Bruton:2023ApJ...949L..40B, Hsiao:2023nix, tang2024jwstnirspec, 
Curtis-Lake:2023NatAs...7..622C, Nakane:2024ApJ...967...28N}. 
We also plot the ionization history of bump models 
with the maximum allowed amplitude $A_{\rm I}$ 
inferred from $\tau_{\rm Planck}$.
{Since these data points are inherently model-dependent
and exhibit partial inconsistencies,
we do not perform a quantitative comparison
with the expected ionization histories.}
Nevertheless, we note that the predicted ionization histories
for the bump models generally fall
within the observational uncertainties,
providing no significantly tighter constraints.


\newpage
\section{Summary and Discussion}
\label{sec:Discussion}
In this work, we analyzed the effects of 
features in the primordial power spectrum expected from particle production during inflation 
on the global 21 cm profile and the ionization history. 
The bump-like primordial features are parameterized by their amplitude  
$A_{\rm I}$ and the scale corresponding to the peak of the  
feature $k_{\rm peak}$.  
Using the semi-numerical simulation code \texttt{21cmFAST} \cite{Mesinger:2011,Murray:2020trn}, we modify the power spectrum and simulate the  
ionization history and the global 21 cm profile  
for various values of $A_{\rm I}$ and $k_{\rm peak}$.  

{The first part of our study is a detailed investigation of the impact of primordial bump-like features on the reionization history and the sky-averaged 21 cm signal. When the background EoR model is fixed, we found that} introducing primordial bump-like features 
can significantly alter the ionization history of the universe  
and the global 21 cm profile
(as shown in figure~\ref{fig:global21_all}).  
The extent of deviation from the  
fiducial model is very sensitive to the scale at which the primordial  
feature peaks.  
{An important finding} is that, for a given EoR model,
there exists a special turnover scale $k^{\rm turn}$, 
which depends on the virial temperature $T_{\rm vir}$ 
(refer to figure~\ref{fig:global_BrightGal}) and   
controls the behaviour of primordial features on the 21~cm signal as follows:
\begin{itemize}
            \item {If 
            $k_{\rm peak}$} is less than the scale $ k^{\rm turn}$, 
            reionization ends later than the fiducial model. 
            Also, the global 21 cm profile was found to shift toward low-redshift. 
            An increase in the amplitude of the primordial feature 
            enhances this effect. 
            %
            \item {If 
            $k_{\rm peak}$ is greater than 
            $ k^{\rm turn}$}, 
            reionization ends earlier than the fiducial model and 
            the global profile of 21 cm shifts toward the high-redshift side.  
            \item When the primordial features 
            peak around $ k^{\rm turn}$, 
            the reionization history and the global 21 cm profile
            were almost unchanged compared to the fiducial model, i.e., the effects of the bump-like features vanish. 
\end{itemize}
We found that the above-mentioned effects {can be traced back to} 
how  the bump-like features influence structure formation  
at different redshifts.  
In section~\ref{subsubsec:HMF}, we provide a detailed  
investigation of the effects of primordial features on the  
Halo Mass Function (HMF).  
Introducing bump-like features in the primordial  
power spectrum affects the  
variance of the smoothed density fluctuations,  
thereby impacting the HMF.  
In particular, the contribution of primordial 
features to the HMF in the low-mass regime {($M_h > M_{\rm min}$)}   
is important to understand the effects on structure formation. 
Our {results show} 
that if the primordial features  
have $k_{\rm peak} < k^{\rm turn}$,  
the tail of the feature affects the HMF 
in the low-mass regime such that 
the number density of halos is less than that of the  
fiducial model, leading to  delayed reionization  
and a shift in the global 21 cm profile toward lower redshifts.  
Conversely, if $k_{\rm peak} > k^{\rm turn}$,  
the peak of the feature affects the low-mass range,  
increasing the number density of halos  
compared to the fiducial model.  
As a result, a larger number of low-mass galaxies  
participate in structure formation,  
causing reionization to complete earlier.  
At $k_{\rm peak} \simeq k^{\rm turn}$,
the tail of the bump coincides with the fiducial model
in the low-mass range of HMF, 
causing negligible effects on the global 21 cm signal.  

The second part of our investigation compares the effects of primordial features  
with the effects of three astrophysical parameters: ionizing efficiency $\zeta$,  virial temperature $T_{\rm vir}$, and  
X-ray luminosity $L_X$.  
Our main finding is that primordial bump-like features  
exhibit 
{degeneracy between $T_{\rm vir}$ and $k_{\rm peak}$, with these two parameters being anti-correlated.}  
{On the other hand,} $\zeta$ and $L_X$ show  
distinguishable effects from the primordial features. 

{Lastly,} by comparing  
the expected optical depth to reionization $\tau_e$ 
for the bump models 
with the observed value from {\it Planck},  
we derived upper limits on the amplitude of the features  
in the {$k$ range [$0.1, 100$] ${\rm Mpc}^{-1}$, assuming a fiducial EoR model.}  
Our results in figure~\ref{fig:tau_values}  
indicate that the scale $k^{\rm turn}$ remains unconstrained, 
while the tightest upper limit {is obtained for the scales} 
where the features have the highest impact 
on the HMF in the low-mass range, 
i.e., for $k_{\rm peak} \sim 7\, {\rm Mpc}^{-1}$.  
Additionally, values of $k_{\rm peak} > 70\, {\rm Mpc}^{-1}$  
are unconstrained as the effects of primordial features  
diminish in the HMF.   
{However, the constraints are sensitive to the underlying astrophysical parameters.}


\vskip 2mm
\noindent{\it Future direction:}  
We note that the effects of primordial bump-like features  
on the 21 cm signal are {very} 
sensitive to the scale  at which they peak.  
The {effect} 
of $k_{\rm peak}$ on the ionization history and the global 
21 cm profile can be mimicked up to a certain extent by tuning the 
parameter $T_{\rm vir}$. 
{This degeneracy and the fact that primordial bump models are unconstrained if $k_{\rm peak}=k^{\rm turn}$
are limitations when we try to constrain such models using the global 21 cm profile alone.} 
Therefore,  combining other observables, 
such as 21 cm power spectra, 
{higher-order statistics}, 
and morphological statistics, 
will be necessary to break this degeneracy.
A detailed analysis of the effects of bump-like features
on these observables will be presented in an upcoming work.

Besides the bump-like primordial features discussed in this work,  
other types of features, such as oscillatory, sharp rises,  
or dips in the primordial power spectrum, may leave unique  
signatures on the 21 cm signal.  
Since the variance of the density fluctuations involves  
the convolution of the power spectrum with a window function,  
the shape of the feature and the range of mass scales  
they affect will be important for estimating their HMF.  
Such a detailed study is worth investigating to check  
if future observations can distinguish the signatures of different  
inflationary models.  This will be carried out as part of future work.  

\section*{Acknowledgements}
\label{sec:ack}
We thank the anonymous referee for the helpful comments/suggestions. 
The computational work in this paper was carried out using 
the NOVA cluster at IIA, Bangalore. 
SSN is supported by 
the SERB - National Post Doctoral Fellowship 
by 
the 
Anusandhan National Research Foundation, 
Government of India,
under the project file number PDF/2023/001469.
KF acknowledges Manipal Centre for Natural Sciences, Centre of
Excellence, Manipal Academy of Higher Education (MAHE) for facilities and support.

\appendix
\section{Effects of primordial features on the quantities 
contributing to the brightness temperature of the 21 cm signal}
\label{app:}
In this section, we discuss the contributions of primordial features to density fluctuations $\delta_{\rm nl}$, 
spin temperature $T_S$ and the neutral fraction of hydrogen $x_{\rm HI}$, individually. 
The differential brightness temperature of the 21 cm signal is given by 
\begin{equation}
    \delta T_{b} \approx
    27\, x_{\rm HI}\, 
    (1+ \delta_{\rm nl}) \, 
    \left(\frac{H}{{dv_r/dr} + H}\right)
    \left(1 - \frac{T_\gamma}{T_S}\right)
    \left(\frac{1+z}{10} \frac{0.15}{\Omega_M h^2}\right)^{1/2}
    \left(\frac{\Omega_b h^2}{0.023}\right)\,
    {\rm mK}.
    \label{eq:delta_Tb_approx}
\end{equation}
\begin{figure}[tbp]
    \centering
    (a)
    \begin{subfigure}{\textwidth}
    \includegraphics[width=\textwidth]{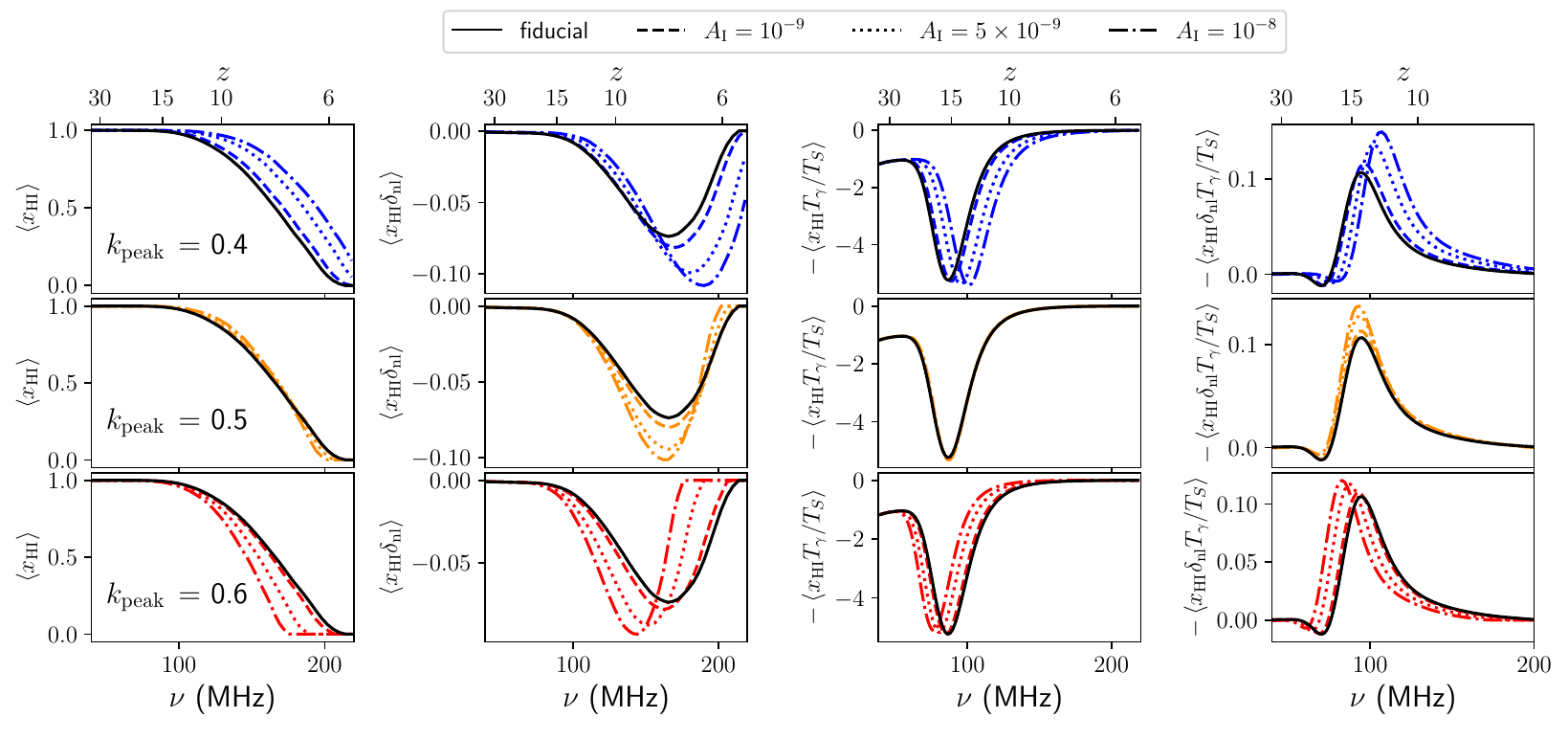}
    \end{subfigure}
    \hfill
    \\(b)\\
    \begin{subfigure}{\textwidth}
        \includegraphics[width=\textwidth]{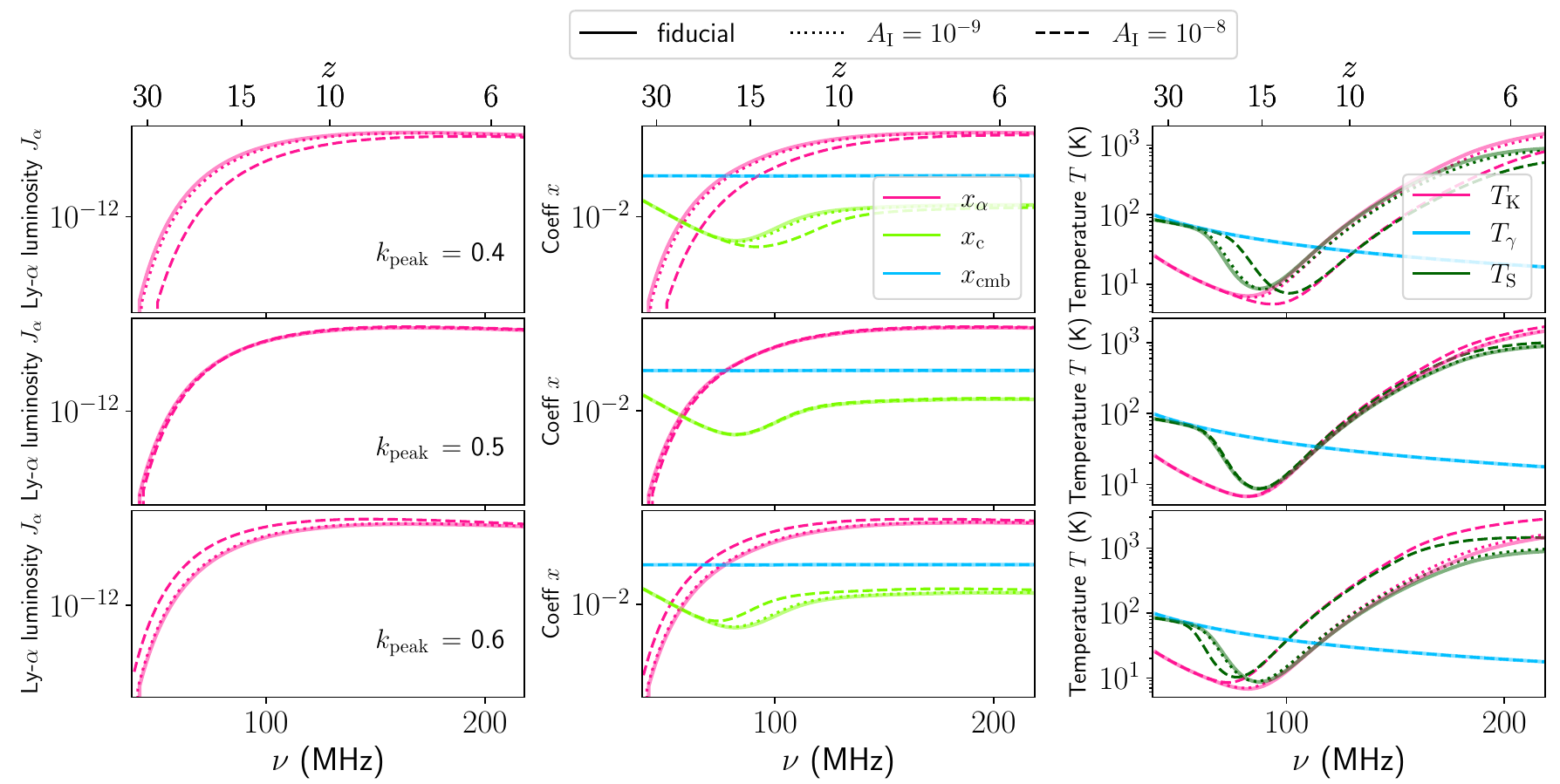}
    \end{subfigure}
\caption{(a) The average quantities appearing on the right-hand side of eq.~\eqref{eq:av_delta_Tb} for the fiducial 
    and bump models. The top, middle, and bottom panels correspond to 
    $k_{\rm peak}$ values of 0.4, 0.5, and 0.6, respectively.
    The value of $A_{\rm I}$ are shown by different line styles. 
    (b)  The three columns show the total Ly-$\alpha$ luminosity, coupling coefficients, and evolution of the average temperatures for the fiducial 
    and bump models. 
}
 \label{fig:app}
\end{figure}
To understand how the quantities $\delta_{\rm nl}$, 
$T_S$ and $x_{\rm HI}$ affect the sky-averaged 21~cm signal, 
we take the average of the 
above equation and obtain the following:
\begin{equation}
  \bigg\langle \delta T_{b} \bigg\rangle=
    C(z)\left[ \bigg\langle x_{\rm HI}\bigg\rangle
    +\bigg\langle x_{\rm HI} \delta_{\rm nl}\bigg\rangle  
    -  \bigg\langle x_{\rm HI} \,\frac{T_{\gamma}}{T_S}\bigg\rangle 
    - \bigg\langle x_{\rm HI} \delta_{\rm nl} \,
    \frac{T_{\gamma}}{T_S}\bigg\rangle \right],
      \label{eq:av_delta_Tb}
\end{equation}
where 
\begin{equation}
C(z) = 27 \left(\frac{\Omega_b h^2}{0.023}\right)  
\left(\frac{0.15}{\Omega_M h^2}\right)^{1/2} 
\left(\frac{1}{10} \right)^{1/2} 
\left(1+z \right)^{1/2}\,.
\end{equation}
The averaged quantities on the right hand side of the above equation are plotted for three 
values of $k_{\rm peak}$ in the three rows of figure~\ref{fig:app}(a).
As previously noticed, the effects on all the quantities are quite opposite for 
$k_{\rm peak} < k^{\rm turn}$
and $k_{\rm peak} > k^{\rm turn}$. 
The effects of the amplitude parameter are shown by different linestyles. 
Focusing on the middle row, 
where $k_{\rm peak} \sim k^{\rm turn}$, 
the contributions of the bump-like features 
are prominent in the quantities $\left< x_{\rm HI} \delta_{\rm nl} \right>$ 
and $- \left< x_{\rm HI} \delta_{\rm nl} \,\frac{T_{\gamma}}{T_S} \right>$ 
due to the density fluctuations, $\delta_{\rm nl}$. 
However, the other two quantities, $\left< x_{\rm HI}\right>$ and 
$- \left< x_{\rm HI} \,\frac{T_{\gamma}}{T_S} \right>$, 
which are independent of $\delta_{\rm nl}$, 
are relatively insensitive to variation of $A_I$ for $k_{\rm peak}\sim 0.5$. 
In addition, the spin temperature $T_S$ is calculated as follows:
\begin{equation}
Ts^{-1} = 
\frac
{x_{\rm cmb} T_\gamma^{-1} + 
x_\alpha T_K^{-1} 
+ x_c T_K^{-1}}
{x_{\rm cmb} + x_\alpha + x_c}\,,
\label{eq:spinT}
\end{equation}
where
$x_{\rm cmb}$, $x_\alpha$ and $x_c$
are the coupling coefficients due to interactions with the 
CMB photons, Ly-$\alpha$ photons and 
collisions, respectively.
$T_K$ and $T_\gamma$ are the kinetic temperature and 
CMB temperature, respectively. 
We plot all these quantities for the three cases 
of bump models in figure~\ref{fig:app}(b).
The redshift evolution of the 
Ly-$\alpha$ luminosity is also plotted in the first column. 
Consistent with the previous arguments, 
the distinct behaviour of bump-like features 
for $k_{\rm peak}$ above and below $k^{\rm turn}$ and 
negligible effects for $k_{\rm peak} \sim k^{\rm turn}$ 
is apparent in all of the plots. 
The major quantity that influences these parameters is the 
collapsed fraction, $f_{\rm coll}$, which depends on
the halo mass function, as discussed in section~\ref{subsubsec:HMF} (also see appendix~\ref{subsec:fcoll} below). 
If $k_{\rm peak} < k^{\rm turn}$ ($ > k^{\rm turn}$), 
the structure formation is driven by a low (high) number 
of low-mass galaxies compared to the fiducial model, 
which is also evident from the suppressed (enhanced)
Ly-$\alpha$ luminosity at every redshift. 
\section{Relation between the peak of the primordial features and mass of the halos}
\label{app:kpeak_m}
\begin{figure}[tbp]
    \centering
    \includegraphics[width=0.5\linewidth]{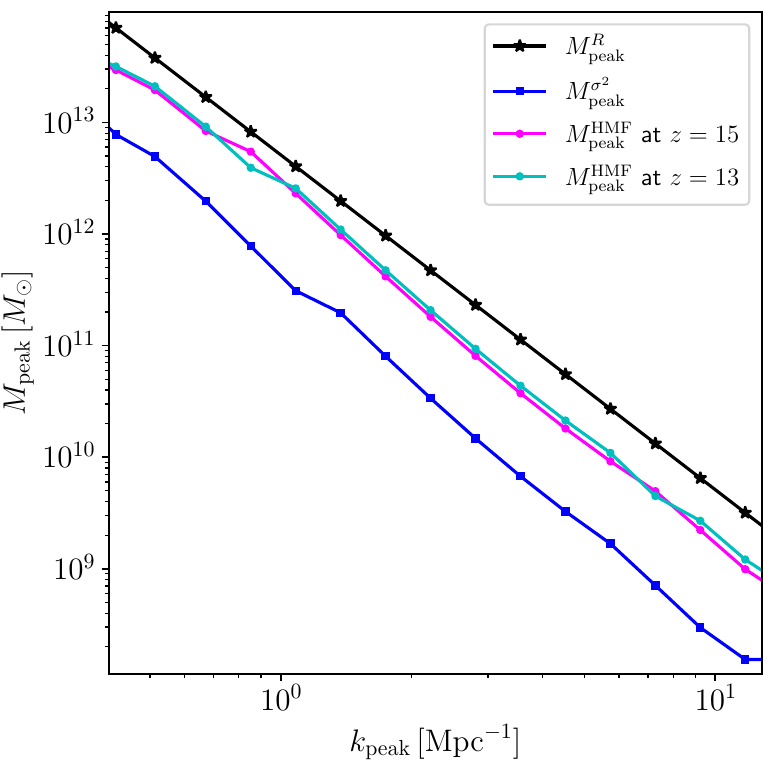}
    \caption{The relation between $k_{\rm peak}$ and $M_{\rm peak}^R$  
calculated using the window function is  
plotted in black. The values of $M_{\rm peak}^{\sigma^2}$, i.e., mass scale   
associated with $k_{\rm peak}$ derived from $\sigma^2(M, z=0)$
in figure~\ref{fig:ST_HMF_bumps}, are shown in blue.  
The magenta and cyan lines represent $M_{\rm peak}^{\rm HMF}$
at redshifts $z=15$  
and $z=13$, respectively.   }
    \label{fig:k_M}
\end{figure}
As discussed in section~\ref{subsubsec:HMF}, the peak of the primordial features, 
$k_{\rm peak}$, can be linked to a halo mass scale $M_{\rm peak}^R$ by comparing the spatial scale 
$(2\pi/k_{\rm peak})$ to the radius of the top-hat window function. 
This relationship between $k_{\rm peak}$ and $M_{\rm peak}$ is shown in Figure~\ref{fig:k_M}.

In addition, the variance of density fluctuations, $\sigma^2(M, z=0)$, 
and the HMF for the bump models shown in Figure~\ref{fig:ST_HMF_bumps} 
indicate that as $k_{\rm peak}$ increases, 
the mass scales associated with the peak of the profile, i.e.,  $M_{\rm peak}^{\sigma^2}$ 
and $M_{\rm peak}^{\rm HMF}$, 
shifts to lower values. 
The corresponding $k_{\rm peak}$ and $M_{\rm peak}$ values 
are also plotted in Figure~\ref{fig:k_M} for comparison.

We note that the $M_{\rm peak}^{\sigma^2}$ and  $M_{\rm peak}^{\rm HMF}$ 
exhibit a similar anti-correlation with $k_{\rm peak}$, as expected, 
with a shift in amplitude.

\section{Collapsed fraction of the halos}
\label{subsec:fcoll}
In this section, we show the effects 
of bump-like primordial features on the 
collapsed fraction of halos, $f_{\rm coll}$. 
The ionization fraction and the spin temperature 
depend on $f_{\rm coll}$ and, therefore, affect the 
differential brightness temperature of the 21 cm line. 

\begin{figure}[tbp]
    \centering
    \includegraphics[width=0.8\textwidth]{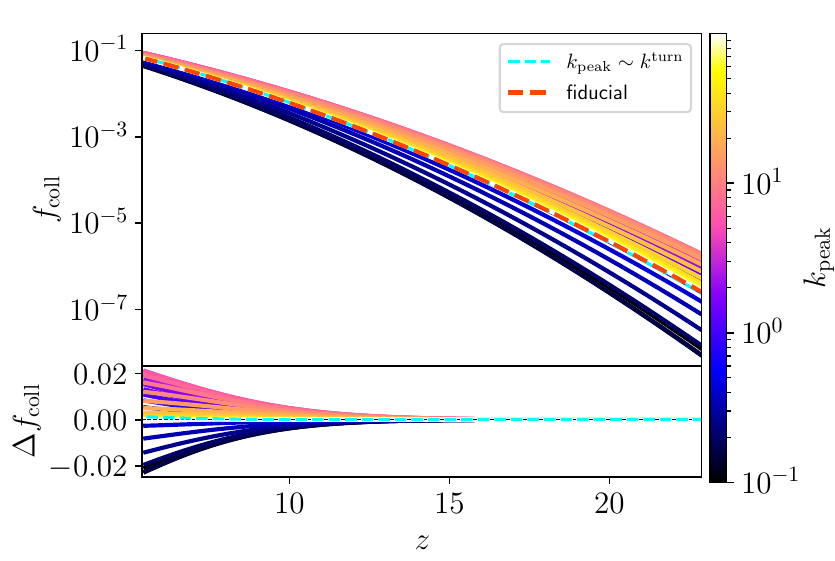}
    \caption{{[Top]} The collapsed fraction $f_{\rm coll}$ for the fiducial model and bump models. 
    {[Bottom]} Difference in the collapsed fraction of the bump models relative to the fiducial model, i.e., $\Delta f_{\rm coll} = f_{\rm coll}^{\rm bump} - 
    f_{\rm coll}^{\rm fid}$.}
    \label{fig:fcoll}
\end{figure}

The HMF calculated from the Sheth - Tormen mass function 
is used to estimate the collapsed fraction of the halos, $f_{\rm coll}$, 
given by the fraction of mass contained in halos with mass greater than 
a threshold $M_{\rm min}$ at redshift $z$. 
The redshift dependence of the collapsed fractions for the primordial features is plotted 
in figure~\ref{fig:fcoll}. 
Following our earlier argument, 
it can be noticed from the figure that, 
at a given redshift $z$, 
a bump-like feature that peaks at 
$k_{\rm peak} < k^{\rm turn} (> k^{\rm turn})[{\rm Mpc}^{-1}]$
shows a lower (higher) collapsed fraction as it
slows down (speeds up) the structure formation, 
which further results in delayed (earlier)
end of reionization. 
For the case of a bump at $k_{\rm peak} \sim k^{\rm turn} = 0.5 {\rm Mpc}^{-1}$, 
the deviation from the fiducial model  
is not significant enough to alter the reionization history 
and the global 21~cm profile. 

\newpage

\def\apj{ApJ}%
\def\mnras{MNRAS}%
\def\aap{A\&A}%
\def\apjl{ApJ}
\def\aj{AJ}
\def\physrep{PhR}
\def\apjs{ApJS}
\def\jcap{JCAP}
\def\pasa{PASA}
\def\pasj{PASJ}
\def\nat{Natur}
\def\apss{Ap\&SS}
\def\araa{ARA\&A}
\def\aaps{A\&AS}
\def\ssr{Space Sci. Rev.}
\def\pasp{PASP}
\def\na{New A}
\def\psj{PSJ}

\bibliography{ref_21cm.bib}
\bibliographystyle{JHEP}

\end{document}